\pdfoutput=1
\documentclass[preprint,authoryear,12pt]{elsarticle}

\usepackage{amsmath,amssymb}
\usepackage{graphicx}
\usepackage{booktabs}
\usepackage{microtype}
\usepackage[colorlinks=true,allcolors=blue]{hyperref}

\journal{Computers \& Geosciences}

\begin{document}
\begin{frontmatter}

\title{Reliability-calibrated deep residual full-waveform inversion using
geometry-invariant physics encoding: synthetic validation and zero-shot
Marmousi-2 testing}

\author[igf]{Deepak Kumar\corref{cor1}}
\ead{deepak.kumar@igf.edu.pl}
\author[allduniv]{Jayant Nath Tripathi}
\ead{jntripathi@allduniv.ac.in}
\author[ngri]{Laxmidhar Behera}
\ead{laxmidhar@ngri.res.in}
\cortext[cor1]{Corresponding author.}
\affiliation[igf]{organization={Institute of Geophysics, Polish Academy of
Sciences}, city={Warsaw}, country={Poland}}
\affiliation[allduniv]{organization={University of Allahabad},
city={Prayagraj}, country={India}}
\affiliation[ngri]{organization={CSIR--National Geophysical Research
Institute}, city={Hyderabad}, country={India}}

\begin{abstract}
Neural networks that map seismic data directly to velocity models tend to
memorize the acquisition geometry they were trained on, and their
uncertainty estimates are rarely trustworthy once the data drift away from
the training distribution. We describe a complete pipeline, small enough to
run on a laptop, that addresses both problems. The network never sees shot
gathers. Instead, variable acquisition geometries are mapped into a fixed
model-space representation computed by classical physics operators: the
starting model, a regularized classical inversion, two misfit-gradient
images, and six illumination and wavenumber-coverage maps obtained from
fast-marching travel times. The learned component is not a replacement for
FWI; it is a calibrated residual corrector applied to this physics-derived
prior. A six-member heterogeneous ensemble with per-pixel variance heads is
calibrated by physics-conditioned conformal prediction, which provides
finite-sample pixel-wise marginal coverage under exchangeability on the
calibration distribution.
For deployment beyond that distribution we introduce a held-out-shot
physics audit: shots that the inversion never used are simulated through
samples of the predictive distribution, and the interval width is rescaled
where the data-space coverage of those shots peaks. No ground truth is
involved. On a corpus of 1000 synthetic models spanning six acquisition
families, the ensemble reduces error by 38\% relative to its classical
prior and transfers to never-seen geometries without measurable
degradation. Applied zero-shot to the full 17\,km Marmousi-2 line, it
lowers the error from 354 to 304\,m/s while raw coverage collapses to
0.42; the audit restores coverage to 0.89--0.91 across eleven corruption
conditions covering noise, wavelet error, and shot decimation, and its
peak height cleanly separates physics mismatch from benign corruptions.
Budget-matched gather-based baselines, with and without explicit source
conditioning, underperform substantially off their training acquisition or
everywhere, respectively. All code, scripts, and checkpoints will be
released and archived on Zenodo.
\end{abstract}


\end{frontmatter}

\section{Introduction}
\label{sec:intro}

Full-waveform inversion (FWI) estimates subsurface velocity by minimizing
the misfit between observed and simulated wavefields
\citep{tarantola1984,virieux2009}. Decades of experience have identified
what it needs to succeed: low frequencies and an adequate starting model to
avoid cycle skipping \citep{bunks1995,sirgue2004}, a reliable source
wavelet \citep{warner2016}, and regularization to stabilize the poorly
illuminated parts of the model \citep{esser2018,aghamiry2019}. Deep
learning promises a shortcut. Networks trained on synthetic pairs can
produce a velocity model in a single forward pass
\citep{araya2018,wu2019,yang2019,adler2021}, and public benchmarks have
made such experiments easy to run at scale \citep{deng2022}. Two obstacles,
however, still separate these results from tools a practitioner would
trust.

The first obstacle is acquisition fragility. A network that consumes shot
gathers encodes the survey layout in its weights, and its accuracy
collapses when the number of shots, the receiver spacing, or the source
changes \citep{kazei2021,ovcharenko2019}. Operator-learning variants
mitigate this by conditioning on source parameters
\citep{zhu2023fourierdeeponet}, but conditioning is not the same thing as
invariance, and we show below that it does not rescue the approach. The
second obstacle is unreliable uncertainty. Modern networks are
systematically overconfident \citep{guo2017}. Ensembles and Bayesian
approximations \citep{lakshminarayanan2017,gal2016,kendall2017} produce
useful relative uncertainty but carry no guarantee, whereas sampling-based
geophysical approaches \citep{fichtner2018,gebraad2020} and linearized
analyses \citep{ely2018,osypov2013} are expensive or restricted in scope.
Conformal prediction \citep{vovk2005,shafer2008,angelopoulos2023} turns any
heuristic uncertainty into intervals with finite-sample guarantees, but the
guarantee is only as good as the calibration distribution, and field data
are always out of distribution.

This paper describes a pipeline built around four ideas, and one framing
sentence states what it is: a physics-encoded deep residual correction and
uncertainty-calibration framework for classical FWI/ADMM priors, not a
replacement for classical inversion. The first idea is a geometry-invariant
physics encoding (GIPE): the network input is a ten-channel model-space
tensor whose shape is independent of shot count, receiver layout, and
source, because the acquisition influences the input only through the
outputs of physics operators (Section~\ref{sec:gipe}). The second is a heterogeneous deep
ensemble with per-pixel variance heads, calibrated by Mondrian conformal
prediction stratified on physical covariates
(Section~\ref{sec:conformal}). The third is a held-out-shot physics audit
that recalibrates the intervals in deployment without any ground truth,
using the wave equation itself as the calibration oracle
(Section~\ref{sec:audit}); the construction can be read as a physics
analogue of conformal risk control \citep{angelopoulos2024}. The fourth is
an evaluation protocol we consider as important as the architecture:
budget-matched baselines, ablations tied to specific claims, and
robustness sweeps over noise, wavelet error, and shot count at two spatial
scales.

The work extends the ADMM-guided physics-informed framework of our
impedance-inversion study \citep{kumar2026} from post-stack impedance to
pre-stack velocity estimation. Every experiment reported here, including 47
complete inversion chains and twenty audits, ran on a single consumer
laptop (Apple M3~Pro, 18\,GB) in roughly 90 hours of compute
(Section~\ref{sec:setup}). We regard this as a feature: the entire
experimental matrix is reproducible without cluster access.

\section{Related work}
\label{sec:related}

Our classical chain is deliberately conventional. It combines multiscale
frequency continuation \citep{bunks1995,sirgue2004,pratt1999} with
adjoint-state gradients \citep{plessix2006} and a
reweighted-$\ell_1$/total-variation prior
\citep{rudin1992,candes2008} solved with the alternating direction method
of multipliers \citep{boyd2011}, following the constrained and extended
formulations explored by \citet{esser2018}, \citet{peters2017},
\citet{aghamiry2019}, \citet{vanleeuwen2013}, and \citet{brossier2009}.

Data-driven inversion covers a broad spectrum: encoder--decoder networks
mapping gathers to models \citep{wu2019,yang2019,araya2018}, operator
learning with explicit source conditioning \citep{zhu2023fourierdeeponet},
generative geological priors \citep{mosser2020}, theory-guided
formulations \citep{sun2020,richardson2018}, and physics-informed neural
networks \citep{raissi2019,waheed2021}. The encoding proposed here is
closest in spirit to approaches that feed physics-derived images rather
than raw data, such as variational inference on subsurface extensions
\citep{yin2024wise}; what GIPE adds is distribution-free calibration and
an explicit test of geometry transfer.

On the uncertainty side, Hamiltonian Monte Carlo FWI
\citep{fichtner2018,gebraad2020} provides posterior samples at
considerable cost, and deep ensembles or Monte Carlo dropout
\citep{lakshminarayanan2017,gal2016} provide cheap but uncalibrated
spread \citep{guo2017}. Conformal methods
\citep{vovk2005,shafer2008,romano2019,angelopoulos2023} supply
finite-sample marginal guarantees, and Mondrian variants restore
per-stratum validity \citep{vovk2012}. Our contribution is the
combination: conformal calibration on synthetics, plus a physics audit
that transports the calibration to out-of-distribution targets without
labels.

\section{Methodology}
\label{sec:method}

\subsection{Forward model and classical chain}
\label{sec:classical}

We work with the two-dimensional constant-density acoustic wave equation,
\begin{equation}
\frac{1}{c^2(\mathbf{x})}\,\frac{\partial^2 u}{\partial t^2}
- \nabla^2 u = s(t)\,\delta(\mathbf{x}-\mathbf{x}_s),
\label{eq:wave}
\end{equation}
where $c$ is the velocity model, $u$ the pressure wavefield, and $s$ the
source wavelet injected at $\mathbf{x}_s$. Equation~\eqref{eq:wave} is
solved with staggered finite differences (Deepwave;
\citealt{richardson2018}) built on PyTorch \citep{paszke2019}, at
8th-order accuracy for the benchmark data and 4th order for the corpus.

Classical FWI minimizes the least-squares misfit
\begin{equation}
J(c) = \tfrac{1}{2}\sum_{s}\bigl\lVert \mathbf{R}\,u_s(c) -
\mathbf{d}_s \bigr\rVert_2^2,
\label{eq:misfit}
\end{equation}
with $\mathbf{R}$ the restriction to receiver locations and
$\mathbf{d}_s$ the observed gather of shot $s$. The gradient
$g=\partial J/\partial c$ is obtained by the adjoint-state method
\citep{plessix2006}. We run frequency continuation over low-pass bands
$f_1<\dots<f_K$ \citep{bunks1995} with a preconditioned descent step,
\begin{equation}
c \;\leftarrow\; \mathrm{clip}\!\left(c - \alpha_k\,
\frac{P[g]}{\max\lvert P[g]\rvert}\right),
\qquad
P[g] = G_\sigma\!\left[\frac{g\odot m_w}
{\bar{g} + \epsilon\,\max\bar{g}}\right],
\label{eq:precond}
\end{equation}
where $m_w$ masks the known water column, $\bar{g}=G_{\Sigma}[\,\lvert
g\rvert\,]$ is a wide Gaussian smoothing of the gradient magnitude acting
as an illumination compensation, $G_\sigma$ is a light smoothing of the
result, $\alpha_k$ is the step length in m/s for band $k$, and the water
cells are reset to their known values after every update. The masking in
Eq.~\eqref{eq:precond} matters in marine settings: without it, the
near-source water-column artifact absorbs the max-normalization and the
update below the seafloor becomes negligible (Section~\ref{sec:crops}).

The classical result is refined with an edge-preserving prior. Writing
$\mathbf{D}=[\mathbf{D}_z;\mathbf{D}_x]$ for stacked first differences, we
solve
\begin{equation}
\min_{c}\; J(c) + \mu \bigl\lVert \mathbf{M}\odot\mathbf{D}c
\bigr\rVert_1
\label{eq:tv}
\end{equation}
by ADMM \citep{boyd2011}, alternating a data-term descent on the augmented
Lagrangian with the auxiliary and dual updates
\begin{equation}
\mathbf{z} \leftarrow
\mathcal{S}_{(\mu/\rho)\mathbf{M}}\!\left(\mathbf{D}c+\mathbf{u}\right),
\qquad
\mathbf{u} \leftarrow \mathbf{u} + \mathbf{D}c - \mathbf{z},
\label{eq:admm}
\end{equation}
where $\mathcal{S}_\tau(x)=\mathrm{sign}(x)\max(\lvert x\rvert-\tau,0)$ is
the soft-threshold operator. The weights are periodically reweighted,
$\mathbf{M} \propto 1/(\lvert\mathbf{z}\rvert+\varepsilon)$ normalized to
unit mean, following \citet{candes2008}, which sharpens genuine
discontinuities while penalizing oscillatory artifacts. The output of
Eqs.~\eqref{eq:precond}--\eqref{eq:admm} is denoted $c_{\mathrm{ADMM}}$
and serves as the prior for the learned stage.

\subsection{Geometry-invariant physics encoding}
\label{sec:gipe}

\begin{figure}[p]
\centering
\includegraphics[width=\textwidth]{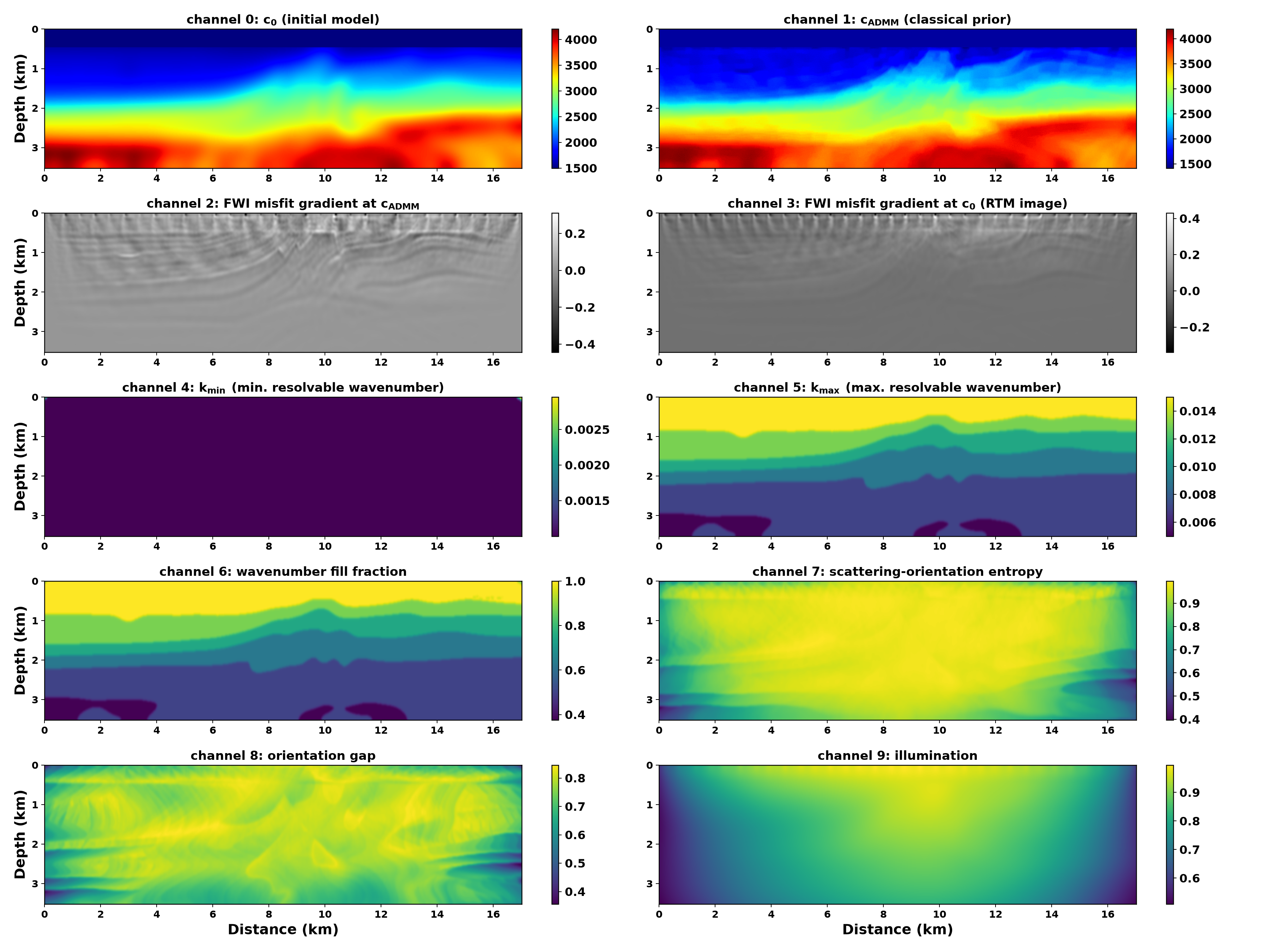}
\caption{The ten-channel geometry-invariant physics encoding of
Eq.~\eqref{eq:gipe} computed for the Marmousi-2 marine line: initial
model, classical ADMM result, misfit-gradient and RTM-like channels, and
the six fast-marching coverage channels of
Eq.~\eqref{eq:wavenumber} (wavenumber bounds and fill, orientation
entropy and gap, illumination).}
\label{fig:gipe}
\end{figure}

The network input is the ten-channel model-space tensor
\begin{equation}
\mathbf{X} = \bigl[\,c_0,\; c_{\mathrm{ADMM}},\;
\tilde{g}_{\mathrm{ADMM}},\; \tilde{g}_{\mathrm{RTM}},\;
\mathbf{C}_k\,\bigr],
\label{eq:gipe}
\end{equation}
where $c_0$ is the smoothed starting model, $\tilde{g}_{\mathrm{ADMM}}$
and $\tilde{g}_{\mathrm{RTM}}$ are max-normalized, lightly smoothed misfit
gradients evaluated at $c_{\mathrm{ADMM}}$ and at $c_0$ (the latter acting
as a reverse-time-migration-like structure image), and $\mathbf{C}_k$ is a
six-channel coverage tensor. To build $\mathbf{C}_k$ we solve the eikonal
equation from every source and every eighth receiver with the fast
marching method \citep{sethian1996}, obtaining travel times $T$ and,
from their normalized gradients, unit ray directions. For a
source--receiver pair
with opening angle $\theta$ at an image point, single scattering
illuminates the local wavenumber
\begin{equation}
k(f) = \frac{2f}{c_0}\cos\frac{\theta}{2},
\label{eq:wavenumber}
\end{equation}
oriented along the bisector of the two ray directions. Accumulating
$k(f)$ over pairs, frequency bands, and orientation bins with geometrical
spreading weights $w \propto [(T_s+t_0)(T_r+t_0)]^{-1}$ yields, per cell:
the minimum and maximum illuminated wavenumber, the wavenumber fill
fraction, the orientation entropy $H=-\sum_b p_b\ln p_b/\ln N_b$, the
orientation gap $1-\max_b p_b$, and the total illumination.
Fig.~\ref{fig:gipe} shows all ten channels for the Marmousi-2 line.

The essential property of Eq.~\eqref{eq:gipe} is architectural. The map
from acquisition to input factors through physics operators whose outputs
live on the model grid, so the input tensor shape is fixed regardless of
shot count, receiver layout, or source, and the network never receives a
representation of the geometry itself. The acquisition still influences
every channel, of course; the claim is not independence but a change of
representation, under which acquisition effects appear as shifts in
channel statistics rather than as changes of input structure. The
augmentation of Section~\ref{sec:setup} was designed to cover those
shifts, and the ablation in Section~\ref{sec:ablations} shows the
residual dependence is mild.

\subsection{Ensemble, predictive variance, and conformal calibration}
\label{sec:conformal}

Six members, spanning three architectures (U-Net, residual CNN, and an
attention U-Net; \citealt{ronneberger2015}) with two seeds each, predict a
residual on the prior together with a per-pixel log-variance,
\begin{equation}
\mu_i = c_{\mathrm{ADMM}} + \lambda\, r_i(\mathbf{X}),
\qquad
\sigma^2_{a,i} = \exp\bigl[\ell_i(\mathbf{X})\bigr],
\label{eq:heads}
\end{equation}
with $\lambda=0.1$ in normalized units. Each member is trained with the
heteroscedastic Gaussian negative log-likelihood
\citep{kendall2017,lakshminarayanan2017}
\begin{equation}
\mathcal{L} = \frac{1}{2N}\sum_{j=1}^{N}\left[
\log \sigma^2_{a}(j) +
\frac{\bigl(v(j)-\mu(j)\bigr)^2}{\sigma^2_{a}(j)}
\right] + \mathcal{R},
\label{eq:nll}
\end{equation}
where $\mathcal{R}$ collects mild gradient-consistency and smoothness
regularizers. The predictive mean and standard deviation combine the
members as
\begin{equation}
\mu = \frac{1}{M}\sum_i \mu_i,
\qquad
\sigma^2 = \underbrace{\frac{1}{M}\sum_i (\mu_i-\mu)^2}_{\text{epistemic}}
\;+\;
\underbrace{\frac{1}{M}\sum_i \sigma^2_{a,i}}_{\text{aleatoric}} .
\label{eq:predictive}
\end{equation}

Raw $\sigma$ is informative but not calibrated. On a held-out calibration
split we compute per-pixel conformal scores and their empirical quantile,
\begin{equation}
s_j = \frac{\lvert v_j - \mu_j\rvert}{\sigma_j},
\qquad
\hat{q}_{1-\alpha} = \mathrm{Quantile}_{\lceil (n+1)(1-\alpha)\rceil/n}
\bigl(\{s_j\}\bigr),
\label{eq:conformal}
\end{equation}
and report intervals $\mu \pm \hat{q}_{1-\alpha}\,\sigma$
\citep{vovk2005,angelopoulos2023}. A precise statement of what is claimed:
all conformal intervals in this paper are empirical pixel-wise marginal
intervals, pooled over cells and calibration instances (or evaluated
within Mondrian strata), valid in the finite-sample sense under
exchangeability of calibration and test instances. We do not claim
simultaneous image-level coverage of the full velocity field, and
neighboring cells are of course strongly correlated. Because marginal validity can hide
systematic under-coverage in poorly illuminated regions, we also compute
Mondrian quantiles \citep{vovk2012} on strata defined by illumination
quartiles and depth halves, a choice we refer to as physics-conditioned
conformal calibration. This is where the coverage channels of
Eq.~\eqref{eq:gipe} earn their keep even after the input ablation: one
cannot stratify on a gradient image, but one can stratify on illumination.

\subsection{Held-out-shot physics audit}
\label{sec:audit}

The guarantee attached to Eq.~\eqref{eq:conformal} holds on the
calibration distribution and lapses under domain shift. The audit
recalibrates without labels. Shots reserved from the entire chain play the
role of a validation set in data space. We draw $M=12$ spatially
correlated fields $\xi_i$ (60\,m correlation length, unit variance) and
form model samples
\begin{equation}
v_i = \mu + \tau\,\frac{\hat{q}_{0.90}\,\sigma}{z_{0.90}}\;\xi_i ,
\qquad z_{0.90}=1.645,
\label{eq:samples}
\end{equation}
so that for $\tau=1$ the marginal spread of the samples matches the
nominal 90\% intervals. Each sample is simulated through the held-out
acquisition with the same physics and wavelet the inversion assumed, and
the observation noise, estimated label-free from pre-first-arrival windows
(no signal can precede the direct water arrival), is added. The data-space
coverage
\begin{equation}
C(\tau) = \frac{1}{\lvert\Omega\rvert}\sum_{(s,r,t)\in\Omega}
\mathbf{1}\Bigl[ d_{srt} \in
\bigl[\,Q_{5}(\tau),\, Q_{95}(\tau)\,\bigr]\Bigr],
\label{eq:datacov}
\end{equation}
is evaluated over the set $\Omega$ of significant samples, where $Q_5$ and
$Q_{95}$ are pointwise percentiles of the simulated ensemble. $C(\tau)$
rises while widening intervals admit more of the data, then decays once
the perturbed wavefields decorrelate from the observations, so its peak
marks the spread that best explains the held-out data. We select
\begin{equation}
\tau_{\mathrm{audit}} = \max\bigl\{\tau : C(\tau) \ge \max_{\tau'}
C(\tau') - \delta \bigr\},
\qquad \delta = 0.005,
\label{eq:plateau}
\end{equation}
the largest value within tolerance of the maximum, preferring
conservative width among indistinguishable optima. The peak location
calibrates; the peak height is diagnostic, because simulating with wrong
physics caps the achievable data-space coverage, so a depressed maximum
flags physics mismatch and its depth tracks severity.

\section{Experimental setup}
\label{sec:setup}

\subsection{Synthetic corpus}

\begin{figure}[p]
\centering
\includegraphics[width=\textwidth]{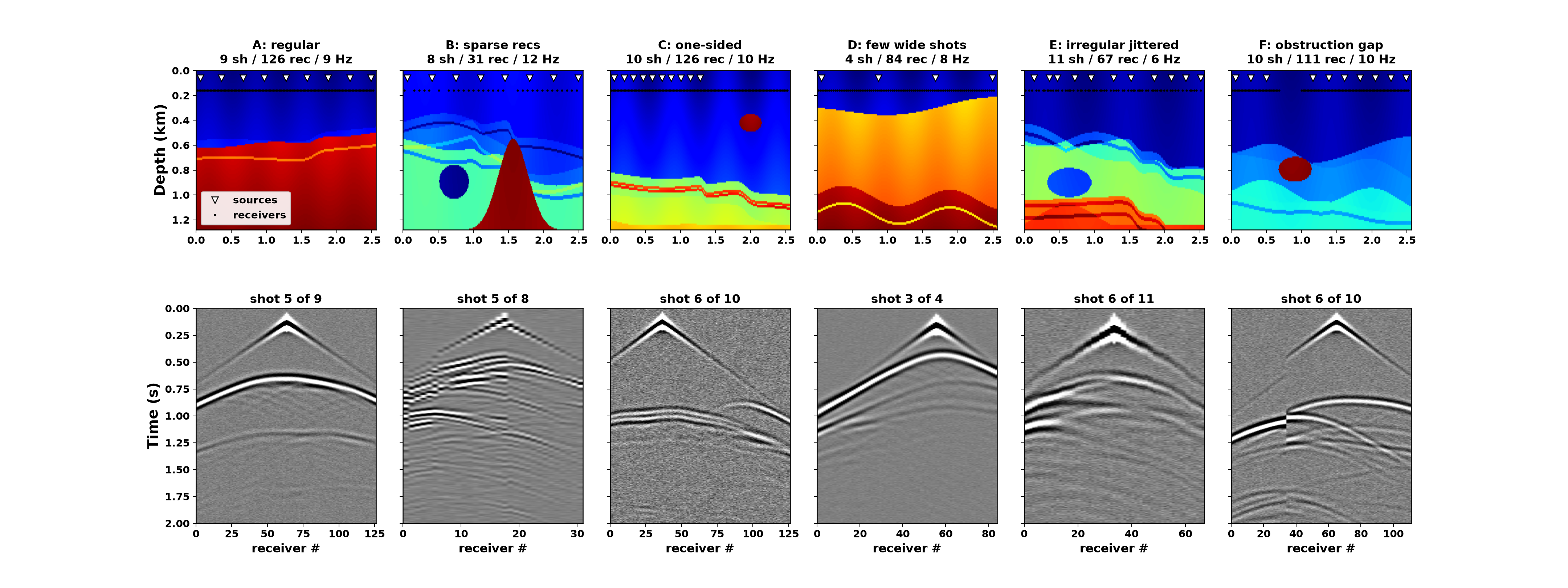}
\caption{Training and evaluation corpus: one instance per acquisition
family, showing the velocity model with its acquisition overlay and one
representative shot gather. Families A to C are seen in training; D to F
appear only at test time.}
\label{fig:corpus}
\end{figure}

We generate 1000 instances of size $128\times256$ cells at 10\,m spacing:
layered backgrounds with folds, up to three faults, lenses, occasional
salt bodies, and lateral velocity trends, split 600/100/100/200 into
train, validation, calibration, and test. Acquisition is drawn per
instance from six families (Fig.~\ref{fig:corpus}): regular spreads with
varying shot count (A), sparse receivers with dropout (B), one-sided
shots (C), few wide shots (D), irregularly jittered shots over non-uniform
receivers (E), and surveys with an obstruction gap (F). Families D, E,
and F never appear in training. The source is a Ricker wavelet with
$f_0\in[6,15]$\,Hz. With probability one half the inversion wavelet is
perturbed relative to the true one, by $\Delta f_0\in[-30,30]\%$ or a
constant-phase rotation up to $90^\circ$, and band-limited noise with
amplitude SNR drawn log-uniformly from $[2,32]$ is added. For every
instance the full classical chain of Section~\ref{sec:classical} and the
tensor of Eq.~\eqref{eq:gipe} are computed. Before corpus generation the
chain was validated on a controlled salt-diapir benchmark
(Supplementary Figs.~S1 and S2).

\subsection{Marmousi-2 benchmark}

\begin{figure}[p]
\centering
\includegraphics[width=\textwidth]{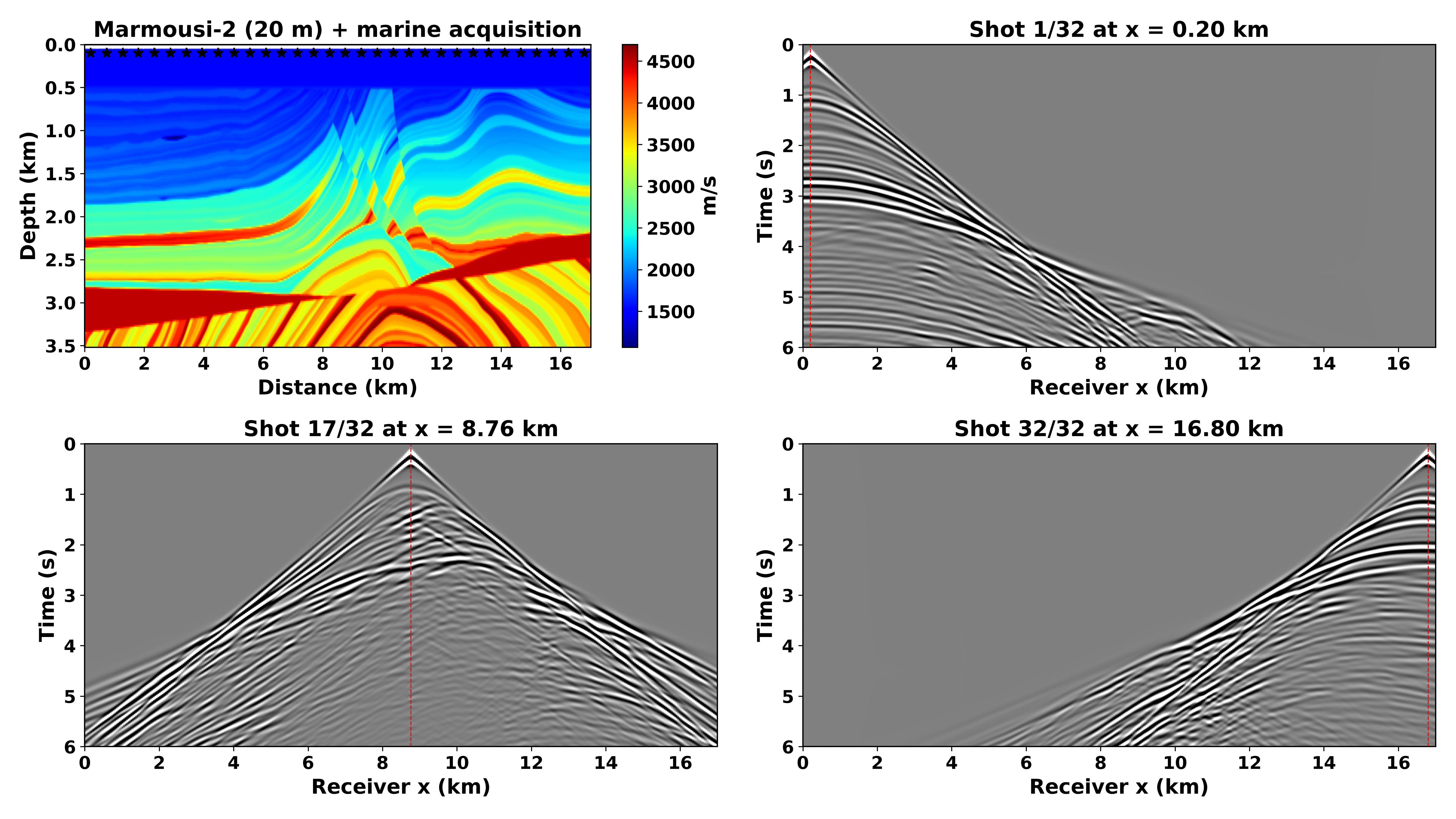}
\caption{Marmousi-2 benchmark data: velocity model with the 32-shot marine
acquisition (stars mark shot positions at the surface), and three
representative shot gathers (6\,s records, free surface, 5\,Hz Ricker
source; the red dashed line marks each source location). Every second
shot is inverted; the remaining shots are reserved for the physics
audit.}
\label{fig:gathers}
\end{figure}

The out-of-distribution target is the full 17\,km Marmousi-2 model
\citep{martin2006} resampled to a 20\,m acoustic grid. The marine survey
comprises 32 shots at the surface, 6\,s records at 2\,ms sampling, a 5\,Hz
Ricker source, a free surface, and receivers every 40\,m
(Fig.~\ref{fig:gathers}); the free surface generates realistic
water-column multiples. The inversion uses every second shot; the odd
shots remain untouched and provide the audit set. The classical chain runs four bands from 3 to 12\,Hz followed
by the ADMM refinement, matching the corpus recipe (per-stage models and
convergence in Supplementary Fig.~S3). No information about
Marmousi, in data or in geometry, enters training or calibration.

\subsection{Implementation and cost}

Wave propagation uses Deepwave 0.0.27 \citep{richardson2018} on PyTorch
2.13 \citep{paszke2019} in float32; eikonal solves use scikit-fmm
\citep{sethian1996}. All experiments ran on a single MacBook~Pro with an
Apple M3~Pro system-on-chip (11 CPU cores: 5 performance, 6 efficiency;
18\,GB unified memory; no discrete GPU). Ensemble members train on the
integrated GPU through the MPS backend; all wave-equation work (corpus
generation, classical chains, audits) runs on CPU.
Table~\ref{tab:cost} itemizes the wall-clock cost of every pipeline
stage; the complete experimental matrix reported in this paper --- corpus,
six ensemble trainings, 47 inversion chains, twenty audits, both
baselines, and both ablations --- amounts to roughly 90 hours of laptop
compute, executed over four days with per-condition checkpointing. The
corpus occupies $\sim$5\,GB on disk, and every random draw is seeded
deterministically by instance or condition tag, so any stage can be
reproduced in isolation.

\begin{table}[t]
\centering
\caption{Wall-clock computational cost of each pipeline stage on the
Apple M3~Pro laptop (float32 throughout; CPU for wave physics, MPS for
training).}
\label{tab:cost}
\footnotesize
\begin{tabular}{@{}lrl@{}}
\toprule
Stage & Cost & Notes\\
\midrule
Corpus instance (model, data, chain, GIPE) & $\sim$2\,min &
$128\times256$, FD order 4\\
Full corpus (1000 instances) & $\sim$33\,h & two parallel shards\\
Ensemble member training & $\sim$10\,min & 100 epochs, MPS\\
Conformal calibration + test evaluation & $\sim$10\,min & 300
instances\\
Full-line FWI (16 shots, 4 bands) & $\sim$62\,min & 20\,m grid, FD
order 8\\
Full-line ADMM refinement & $\sim$25\,min & 8 outer iterations\\
Full-line GIPE channels & $\sim$10\,min & gradients + eikonal\\
Full-line robustness condition (total) & $\sim$1.6\,h & chain +
ensemble inference\\
Full-line audit condition & $\sim$11\,min & reuses stored
$\mu,\sigma$\\
Crop-scale sweep condition & 4--6\,min & $200\times400$ crop\\
Complete experimental matrix & $\sim$90\,h & four days, checkpointed\\
\bottomrule
\end{tabular}
\end{table}

\section{Results}
\label{sec:results}

\subsection{Synthetic corpus: accuracy and calibration}
\label{sec:corpus_results}

\begin{figure}[p]
\centering
\includegraphics[width=\textwidth]{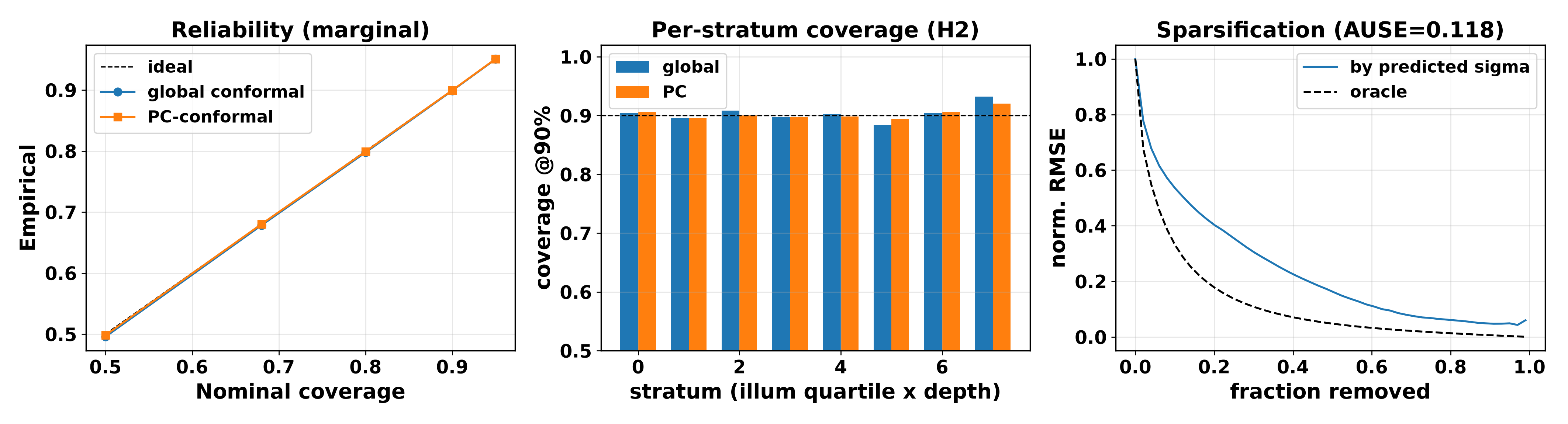}
\caption{Synthetic-corpus evaluation of the calibrated ensemble:
reliability across nominal levels; per-stratum coverage at nominal 0.90
comparing the global conformal quantile with physics-conditioned (PC)
Mondrian calibration over the eight illumination-by-depth strata; and
error sparsification against the oracle ordering (AUSE = 0.118).
Per-family accuracy values are tabulated in Supplementary Table~S3.}
\label{fig:calibration}
\end{figure}

On the 200 test instances the ensemble reaches 128\,m/s RMSE against
205\,m/s for its classical prior, a 38\% reduction
(Fig.~\ref{fig:calibration}). Conformal coverage is essentially exact at
every nominal level. Per-family RMSE spans 111 to 144\,m/s, and the
never-trained families D, E, and F fall inside that range (130, 122, and
137\,m/s), so the zero-shot geometry gap is indistinguishable from the
seen-family spread. The predicted $\sigma$ ranks error well, with a
Spearman correlation of 0.76 and an area under the sparsification error
curve of 0.118. Physics-conditioned calibration lifts the worst
illumination stratum from 0.884 to 0.894 at nominal 0.90.

\subsection{Ablations}
\label{sec:ablations}

Two ablations probe where the transfer actually comes from. Removing the
six coverage channels from the network input changes test RMSE from 128 to
125\,m/s and zero-shot Marmousi from 304 to 305\,m/s; the gradient
channels evidently encode illumination well enough on their own. The
coverage tensor keeps two roles, as Mondrian covariates and as
interpretable diagnostics, and the input can be slimmed to four channels
without loss: the coverage channels are not load-bearing for mean
accuracy, but they are load-bearing for structured calibration, since
one cannot stratify on a gradient image.

The second ablation retrains the ensemble on the same 800 velocity models
but a single fixed acquisition, sixteen regular shots at 10\,Hz and SNR~8
with no wavelet error, and evaluates on the original augmented test set.
The result is 135 versus 128\,m/s, a uniform increase of 5 to 8\,m/s in
every family including the unseen ones, with no geometry-specific gap.
The invariant encoding alone carries the transfer; the acquisition
augmentation acts as a mild regularizer rather than the mechanism. We had
expected the opposite, and report the outcome as it strengthens the
architectural claim.

\subsection{Budget-matched gather-based baselines}
\label{sec:baselines}

\begin{figure}[p]
\centering
\includegraphics[width=\textwidth]{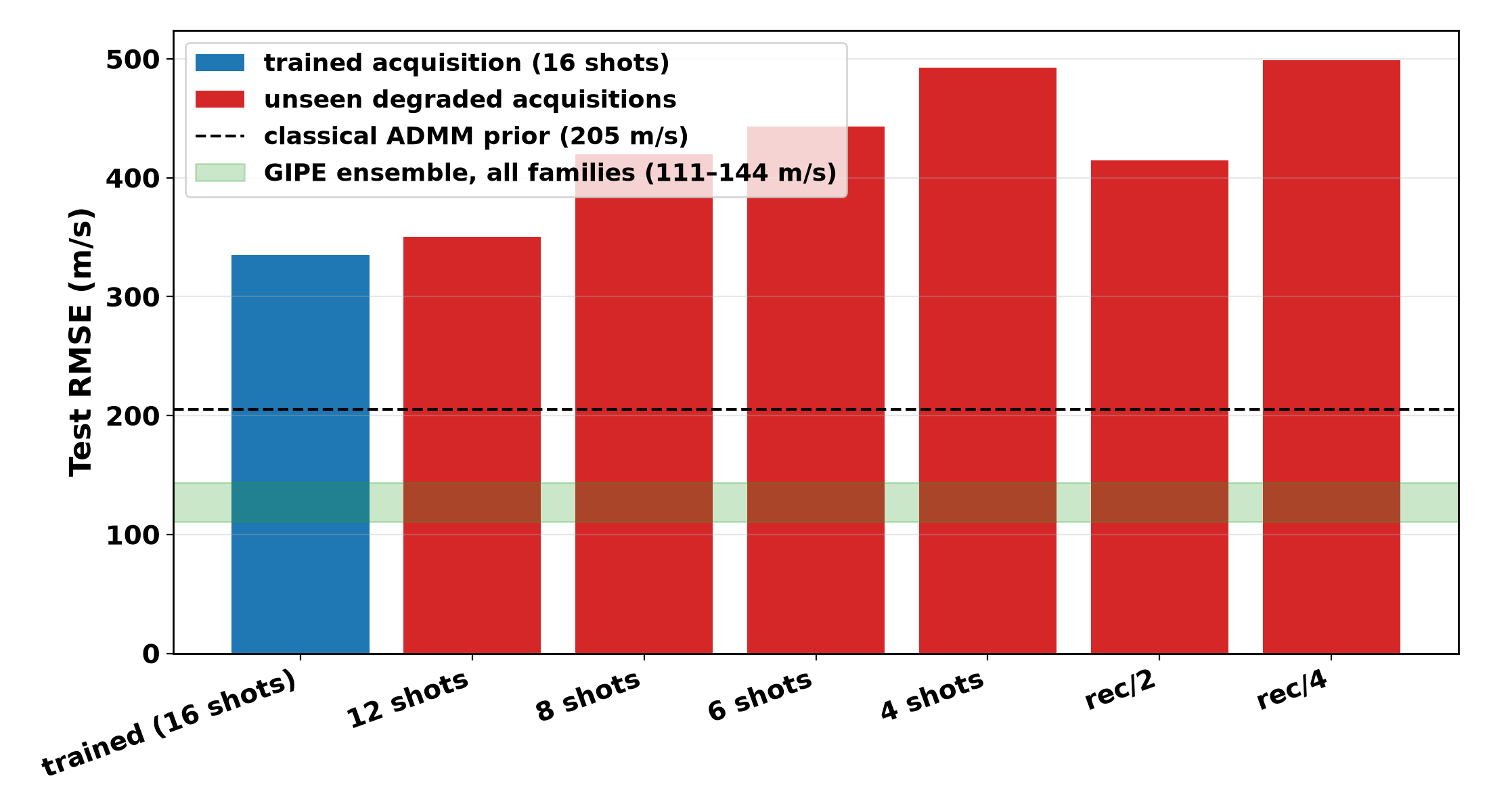}
\caption{Fixed-geometry gather-to-model baseline: the blue bar is the
network evaluated at its own trained acquisition (16 regular shots); red
bars are the same network under unseen shot decimation and receiver
subsampling. The dashed line marks the classical prior and the green
band the GIPE ensemble's range across all families, including unseen
ones.}
\label{fig:collapse}
\end{figure}

\begin{figure}[p]
\centering
\includegraphics[width=\textwidth]{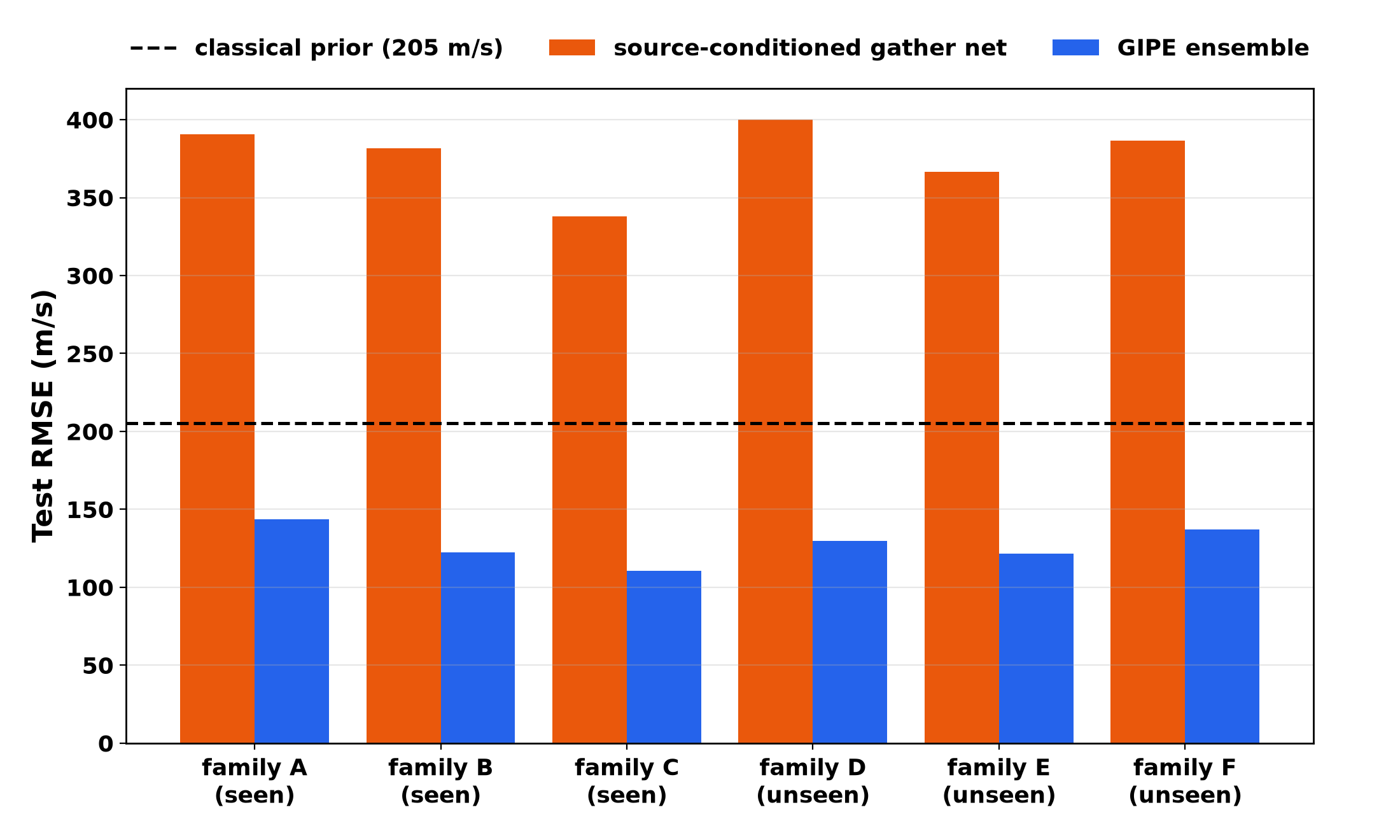}
\caption{Source-conditioned gather network at matched budget: worse than
the classical prior on every family. Explicit conditioning does not
substitute for invariance.}
\label{fig:srccond}
\end{figure}

Two baselines trained on the identical 600-instance budget test the
alternative philosophy (Figs.~\ref{fig:collapse} and \ref{fig:srccond}).
An InversionNet-style network \citep{wu2019} trained at one fixed
acquisition reaches 335\,m/s at that acquisition, already worse than the
classical prior, and degrades to about 495\,m/s when shots are decimated
or receivers subsampled. A source-conditioned variant in the spirit of
Fourier-DeepONet \citep{zhu2023fourierdeeponet}, with a per-shot encoder,
FiLM conditioning on shot position, source frequency, phase, shot count
and receiver density, and 35M parameters, trains across the full augmented
geometry distribution and still lands at 377\,m/s overall, worse than the
classical prior on every family, seen or unseen. At the matched 600-instance
training budget used here, the tested gather-based baselines
underperform the physics-encoded model substantially, whether the
geometry is implicit or explicitly conditioned; we do not claim that
gather-based learning fails in general, only that its sample complexity
was not affordable in this regime. Both baselines used the same
augmentation, output grid, validation criterion, and best-checkpoint
selection as the ensemble; their hyperparameters are listed in the
supplementary material.

\subsection{Zero-shot Marmousi-2 and the audit}
\label{sec:marmousi}

\begin{figure}[p]
\centering
\includegraphics[width=\textwidth]{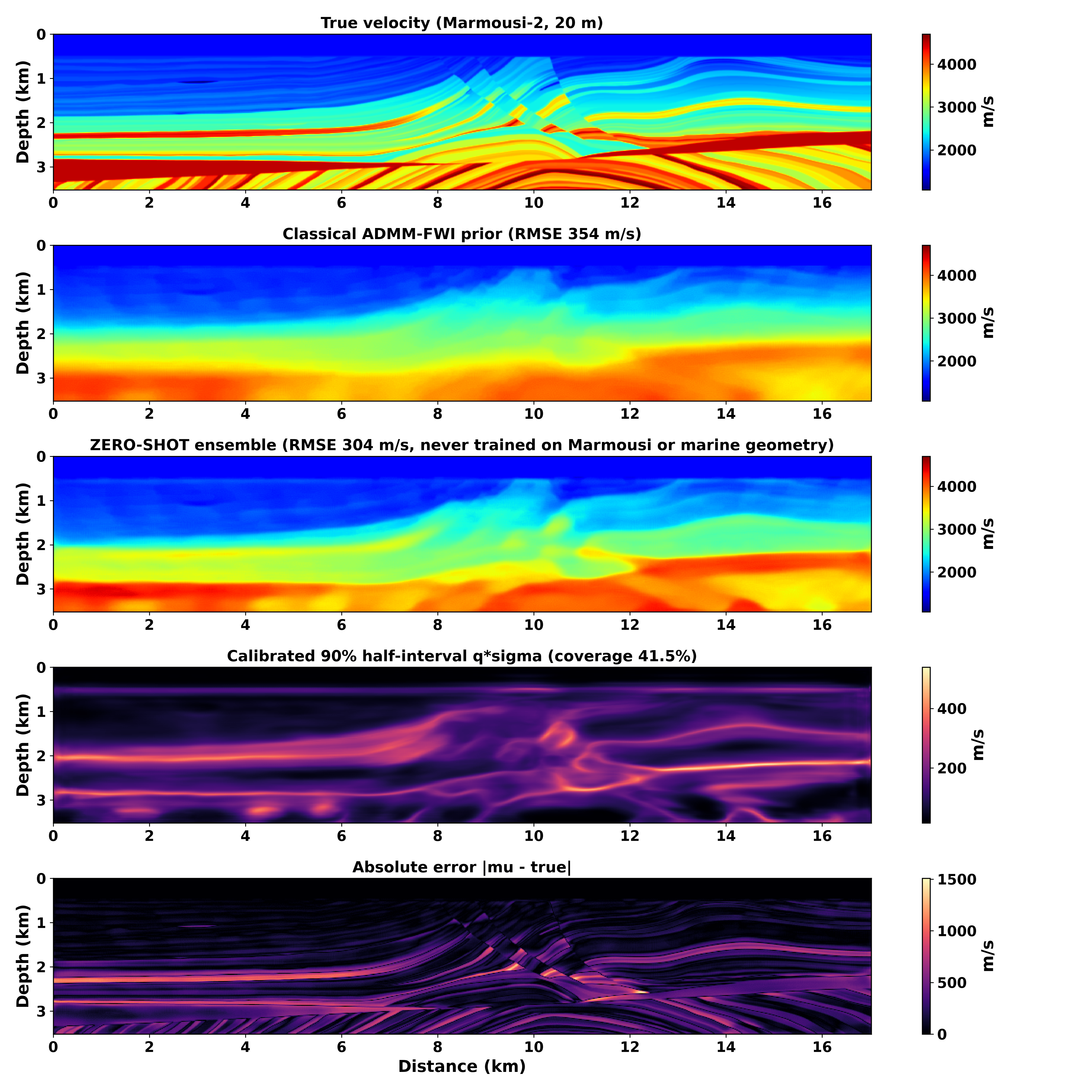}
\caption{Zero-shot application to the full 17\,km Marmousi-2 line: true
model, classical ADMM prior (354\,m/s), ensemble mean (303\,m/s),
calibrated 90\% half-interval, and absolute error.}
\label{fig:zeroshot}
\end{figure}

\begin{figure}[p]
\centering
\includegraphics[width=\textwidth]{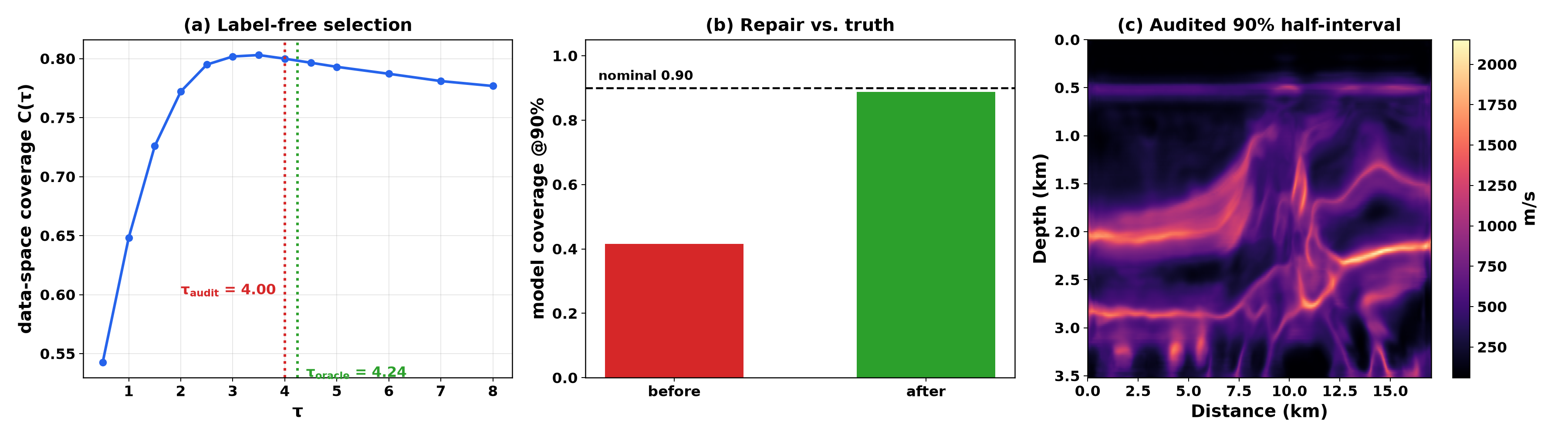}
\caption{Held-out-shot physics audit on zero-shot Marmousi-2: data-space
coverage $C(\tau)$ with the audit and oracle selections, model-space
coverage before and after repair, and the audited half-interval map.}
\label{fig:audit}
\end{figure}

Applied zero-shot to the full line (Fig.~\ref{fig:zeroshot}), the
ensemble improves the classical prior from 354 to 304\,m/s (303 in the 16-shot spine re-run of Table~\ref{tab:spine}). As a
deliberately non-blind upper bound we also fine-tune each member briefly
on the true Marmousi model itself, reaching 268\,m/s
(Supplementary Fig.~S4); this transfer experiment quantifies headroom and
is kept strictly separate from the zero-shot claim. Even with supervision
the gain saturates,
because the residual formulation of Eq.~\eqref{eq:heads} inherits the
prior's reach and cannot rebuild the deep section where the prior carries
no information. Raw conformal coverage collapses from its nominal 0.90 to
0.415, the expected domain-shift failure. The audit, using four held-out
shots and Eqs.~\eqref{eq:samples}--\eqref{eq:plateau}, selects
$\tau_{\mathrm{audit}}=4.0$ against an oracle value of 4.24 and repairs
model-space coverage to 0.888 (Fig.~\ref{fig:audit}).

A harsher test induces a genuine physics mismatch by generating elastic
data and inverting with acoustic operators (Fig.~\ref{fig:elastic};
gather comparison in Supplementary Fig.~S5). This experiment is not an
elastic inversion benchmark; it is a controlled physics-mismatch stress
test of the calibration machinery. The classical chain is
neutralized completely, with FWI and ADMM ending at the initial model,
yet the ensemble still improves the prior from 368 to 328\,m/s. Coverage
breaks to 0.38; the audit repairs it to 0.83, and its peak height drops
to 0.71 against the 0.80 ceiling of the matched-physics case, flagging
the mismatch and its severity without access to any truth
(Supplementary Fig.~S6).

\subsection{Robustness on the full line}
\label{sec:sweeps}

\begin{figure}[p]
\centering
\includegraphics[width=\textwidth]{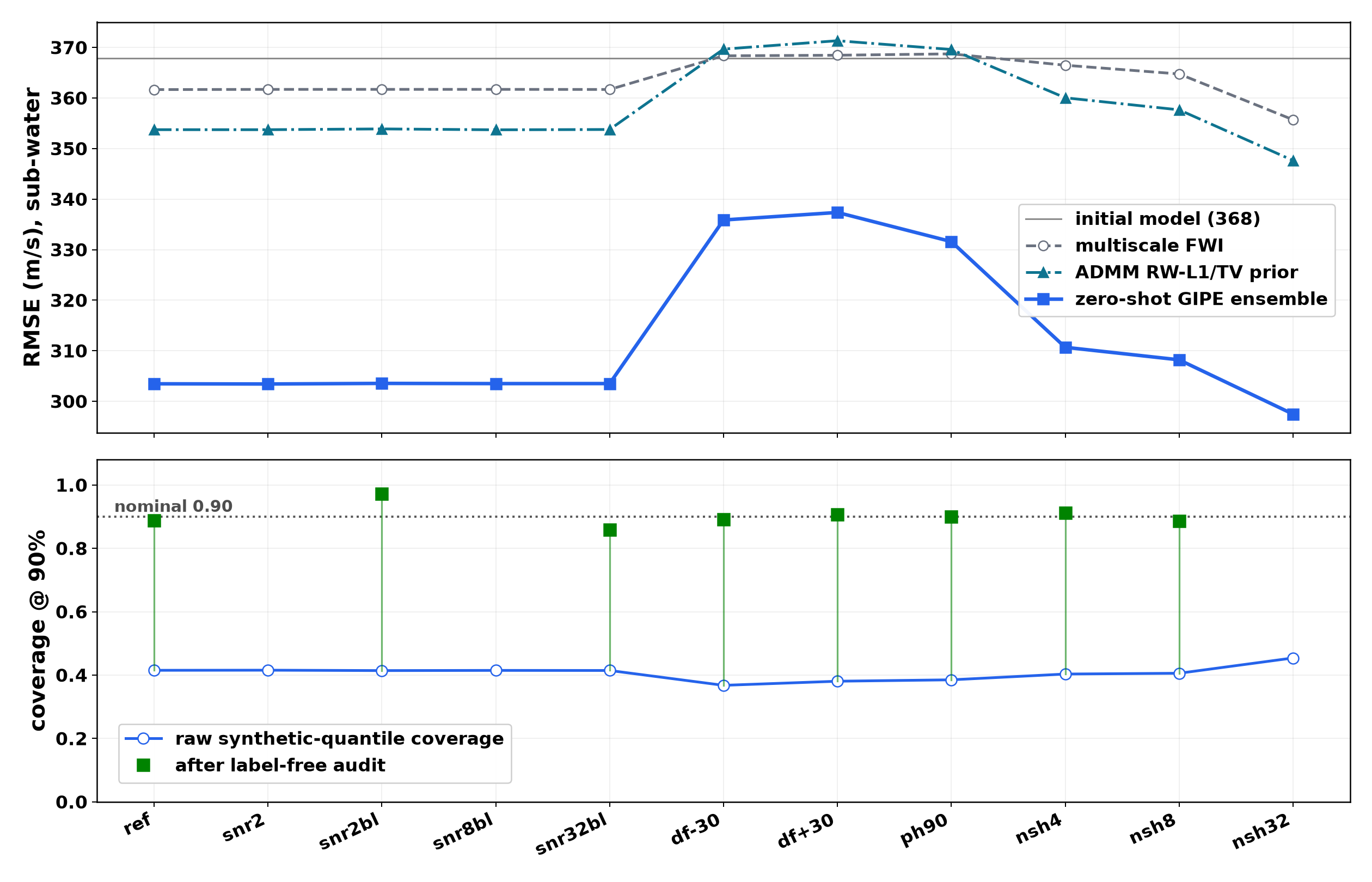}
\caption{Full-line robustness spine: accuracy of FWI, the ADMM prior, and
the zero-shot ensemble across noise, wavelet-error, and shot-count
conditions (top); raw conformal coverage and its label-free audit repair
against the nominal 0.90 (bottom).}
\label{fig:spine}
\end{figure}

Table~\ref{tab:spine} and Fig.~\ref{fig:spine} summarize eleven full-line
conditions. Three observations stand out. Additive noise, whether white or
band-limited, leaves every stage essentially unchanged from SNR~2 to 32;
stacking over sixteen shots and six-second records absorbs it at this
budget. Wavelet error is the axis that kills classical FWI: a 30\%
frequency error or a $90^\circ$ phase rotation drives both FWI and the
ADMM refinement to or slightly above the initial model, yet the ensemble
retains a 9 to 10\% improvement over that same initial model with a dead
prior, reproducing the elastic-mismatch pattern on a second physics-error
axis. Shot count, finally, saturates gently: 4, 8, 16, and 32 shots give
311, 308, 303, and 297\,m/s, so two-thirds of the acquisition can be
discarded at a cost of 8\,m/s.

The audit rows of Table~\ref{tab:spine} close the loop. Across eight
audited conditions the selected $\tau$ falls within 0.25 of the oracle in
seven; repaired coverage lies between 0.886 and 0.912 in six, with a
conservative overshoot to 0.971 only at SNR~2, where the estimated noise
floor rather than model error dominates the predictive band and flattens
$C(\tau)$. Peak height separates the wavelet-mismatch conditions, 0.765 to
0.786, from the benign ones, 0.799 to 0.803. The gap is narrow, so we
report a sensitivity analysis over the audit's hyperparameters (number
of samples, correlation length, held-out-shot choice, plateau tolerance)
in the supplementary material (Table~S6): the repaired coverage stays
within $0.90\pm0.08$ across fourfold changes in sample count and shot
budget, peak height varies by only $\pm0.002$ across held-out-shot
choices at fixed configuration, and the correlation length of the
perturbation fields is the most sensitive knob. A practical rule
follows: compare peak heights only at a fixed audit configuration, and
treat differences exceeding $0.01$ as meaningful, comfortably below the
0.02--0.04 gaps produced by wavelet mismatch here. The raw predictive $\sigma$,
in contrast, is blind to every corruption, sitting at 135 to 137\,m/s
throughout. The network cannot see its own physics mismatch; the audit
can.

\begin{table}[t]
\centering
\caption{Full-line robustness spine on the 17\,km Marmousi-2 line. RMSE in
m/s over the sub-water model; the initial model scores 368\,m/s in every
condition. The audit selects $\tau$ from four held-out shots without
ground truth; the oracle is the inflation that achieves exactly 0.90
against the true model; Peak is the maximum of the data-space coverage
curve $C(\tau)$, the mismatch-severity indicator.}
\label{tab:spine}
\footnotesize
\setlength{\tabcolsep}{3.5pt}
\begin{tabular}{@{}lrrrrccr@{}}
\toprule
Condition & FWI & ADMM & Ens. & Cov.\ raw & $\tau_{\mathrm{audit}}$
(oracle) & Peak & Cov.\ aud.\\
\midrule
reference           & 362 & 354 & 303 & 0.415 & 4.0\,(4.24) & 0.803 & 0.888\\
SNR 2, band-limited & 362 & 354 & 304 & 0.414 & 8.0\,(4.24) & 0.783 & 0.971\\
SNR 32              & 362 & 354 & 304 & 0.414 & 3.5\,(4.25) & 0.799 & 0.859\\
$\Delta f_0=-30\%$  & 368 & 370 & 336 & 0.368 & 5.0\,(5.25) & 0.765 & 0.892\\
$\Delta f_0=+30\%$  & 368 & 371 & 337 & 0.381 & 5.0\,(4.80) & 0.785 & 0.907\\
phase $90^\circ$    & 369 & 370 & 332 & 0.385 & 4.5\,(4.50) & 0.786 & 0.900\\
4 shots             & 366 & 360 & 311 & 0.403 & 4.5\,(4.21) & 0.802 & 0.912\\
8 shots             & 365 & 358 & 308 & 0.406 & 4.0\,(4.28) & 0.803 & 0.886\\
32 shots$^{\dagger}$ & 356 & 348 & 297 & 0.451 & --- & --- & ---\\
\bottomrule
\end{tabular}

\smallskip
{\footnotesize $^{\dagger}$The 32-shot condition uses every shot in the
survey for inversion, so no held-out shots remain and the audit is not
applicable by construction.}
\end{table}

\subsection{Crop-scale sweeps and audit identifiability}
\label{sec:crops}

A denser sweep of 36 conditions on two 4$\times$2\,km Marmousi crops
(Fig.~\ref{fig:crops}; complete values in Supplementary Table~S1)
confirms the same axes at finer resolution and adds
a structure-dependence result: on a faulted thrust crop the ensemble's
gain over the classical chain is a nearly constant $5.3\pm0.3\%$ in every
condition, whereas on a simple layered crop it is neutral. Where the
classical chain already does well, the network has little to add, which is
consistent with the residual formulation inheriting the prior's reach. In
all 36 conditions FWI and its ADMM refinement agree within 1\,m/s; at
these budgets the preconditioned multiscale descent of
Eq.~\eqref{eq:precond} is itself a sufficient regularizer, and a single
classical baseline suffices at crop scale. The ADMM stage earns its
place only at full-line scale, where it contributes 8--12\,m/s over FWI
in every condition of Table~\ref{tab:spine}; we did not retrain the
ensemble on an FWI-only prior, so the learned stage's numbers always
refer to the ADMM prior.

\begin{figure}[p]
\centering
\includegraphics[width=\textwidth]{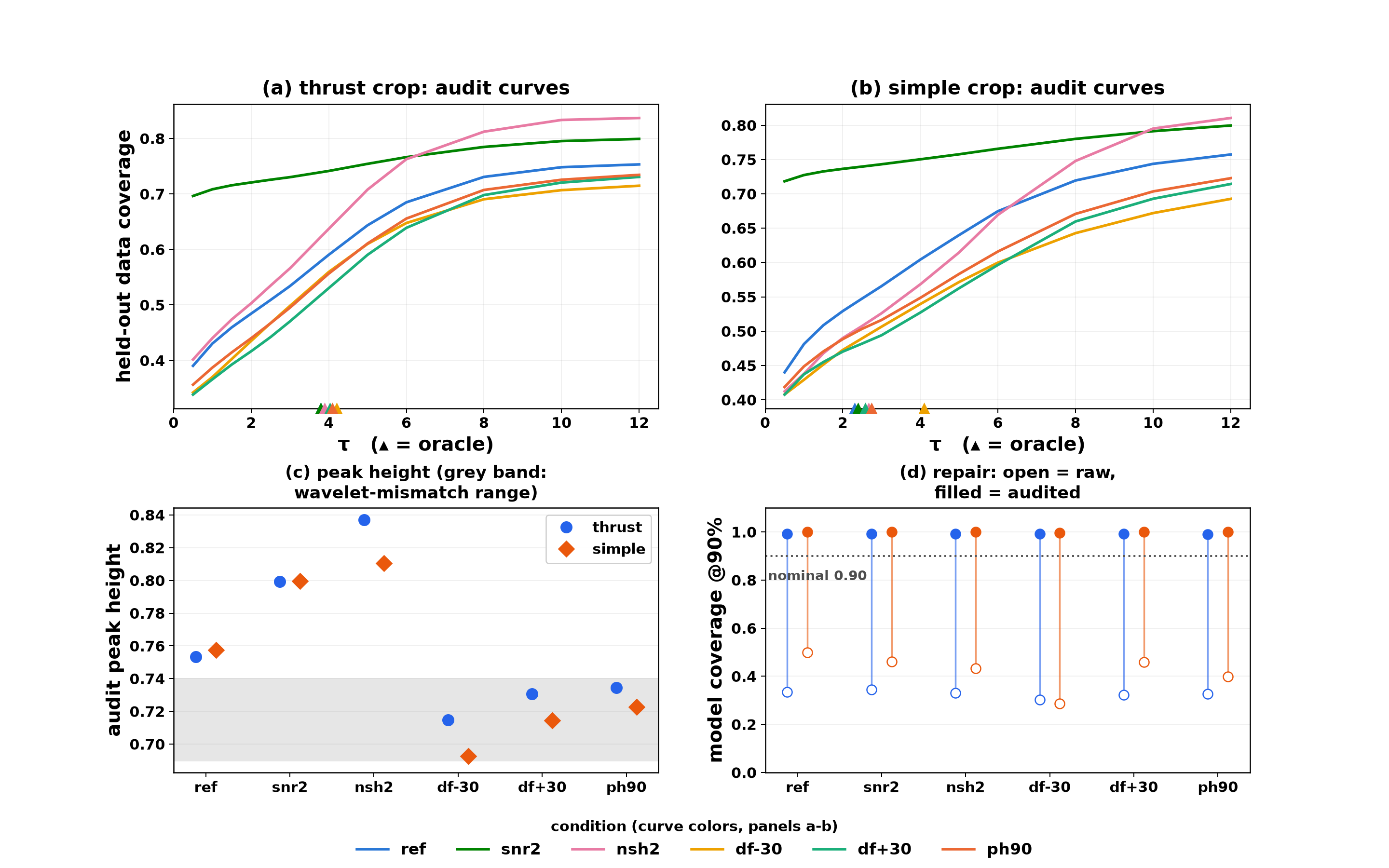}
\caption{Crop-scale audit: $C(\tau)$ curves whose broad plateau reflects
the short-offset identifiability limit, the peak-height severity
indicator, and the conservative coverage repair.}
\label{fig:cropaudit}
\end{figure}

The crop audits expose a scale limitation that we prefer to state plainly
(Fig.~\ref{fig:cropaudit}). With only 4\,km of aperture the curve
$C(\tau)$ rises to a broad, shallow plateau and the selection of
Eq.~\eqref{eq:plateau} runs to the edge of the search grid
(Supplementary Table~S2): perturbed
wavefields accumulate too little travel-time shift over short paths to
decorrelate, so $\tau$ is not identifiable. The failure is conservative,
repairs overshoot toward full coverage and never under-cover, and the
peak height still ranks severity correctly on both crops. Together with
the full-line results this yields a usable identifiability statement: the
audit calibrates when model error dominates the predictive data band over
sufficiently long propagation paths, and degrades to conservative
over-coverage when noise or short offsets starve its sensitivity.

\section{Discussion}
\label{sec:discussion}

Three points deserve emphasis. First, the representation reduces
acquisition dependence by mapping variable geometries into fixed-grid,
physics-derived model-space fields: transfer to unseen geometries then
becomes a property of the representation, and the ablations show that
neither augmentation nor coverage inputs are the load-bearing element. Both gather-based alternatives, implicit and
conditioned, failed at matched budget, which suggests the sample
complexity of learning geometry from data is simply not affordable at
this scale.

Trustworthiness under shift could not be obtained from the network alone.
The predictive $\sigma$ was blind to every corruption we imposed, from
wavelet error to elastic physics, while the wave-equation audit detected
the miscalibration, ranked its severity, and repaired coverage to within
one or two points of nominal in most conditions. We expect this pattern,
heuristic uncertainty plus conformal calibration on synthetics plus a
physics-oracle recalibration in deployment, to transfer to other inverse
problems that possess a reliable forward operator.

It is worth stating plainly why the classical panels of
Figs.~\ref{fig:zeroshot} and \ref{fig:elastic} remain smooth, because the
two cases fail for different reasons. In the acoustic case the prior is
limited by bandwidth and budget, not by implementation: with a 5\,Hz
source and continuation capped at 12\,Hz on the 20\,m grid, the
half-wavelength resolution limit is roughly 100--125\,m in 2.5--3\,km/s
rock \citep{virieux2009}, below which Marmousi-2's thin beds and fine
fault blocks are simply invisible to the data; the sixty preconditioned
iterations then recover the long-to-mid wavelength background, while the
smoothing and illumination compensation of Eq.~\eqref{eq:precond}
deliberately suppress high-wavenumber updates to remain stable at this
budget; and below the diving-wave penetration depth the gradient energy
decays, leaving the deep section near the initial model. In the elastic
case the failure is fundamental rather than budgetary: mode conversions
and elastic amplitudes make the acoustic misfit gradient incoherent with
the true residual, and the chain terminates at the initial model
regardless of iteration count. This asymmetry is what makes the learned
stage's behavior informative: it extracts structure the classical chain
cannot reach in both regimes, and the audit reports honestly how much
that extraction can be trusted.

The limitations are themselves informative, and three deserve concrete
guidance. The audit needs propagation paths long enough for the
perturbed wavefields to decorrelate. A back-of-envelope criterion
follows from requiring the accumulated travel-time perturbation over a
path of length $L$ to reach a quarter period of the dominant frequency
$f$: with velocity $v$ and perturbation scale $\delta v \sim
\hat{q}_{0.90}\bar\sigma$,
\begin{equation}
\frac{L\,\delta v}{v^{2}} \gtrsim \frac{1}{4f}
\quad\Longrightarrow\quad
L \gtrsim \frac{v^{2}}{4 f\, \delta v},
\label{eq:aperture}
\end{equation}
which for $v \approx 2.5$\,km/s, $f \approx 5$\,Hz, and $\delta v
\approx 150$\,m/s gives $L \gtrsim 2$\,km of one-way path. This is
consistent with our observations: identifiability is marginal at 4\,km
of aperture and comfortable at 17\,km. The audit also needs the noise
floor to sit below the model-error signal in the predictive band, a
condition that can be checked in advance from the pre-arrival noise
estimate. Second, the residual formulation caps gains where the prior is
uninformative; mitigations worth exploring include lower-frequency or
tomographic priors, iterating the encode--correct cycle so the corrected
model seeds a new prior, and relaxing the residual scale $\lambda$ with
depth. Third, all experiments are two-dimensional and acoustic at laptop
scale. Nothing in the formulation is dimension-specific: the encoding
channels (models, gradients, fast-marching coverage) are defined
identically on 3-D grids, conformal calibration is dimension-agnostic,
and the audit needs only held-out shots and a forward solver, so the
binding constraint in 3-D elastic settings is the cost of the classical
chain and the predictive simulations rather than the method itself;
frequency-domain or time-domain solvers with checkpointing, and fewer
audit samples at coarser grids, are the obvious levers. The natural next step is a
real marine line, for instance the open Viking Graben dataset, where well
logs would provide sparse ground truth to verify interval coverage at the
boreholes; we leave this to a dedicated study.

\section{Conclusions}

We have described and validated an acquisition-agnostic, calibrated deep
FWI pipeline whose uncertainty statements survive contact with
distribution shift. A physics-computed encoding transfers across six
acquisition families with no measurable gap and does not rely on
augmentation for that ability. Conformal calibration is exact on
distribution, and a label-free held-out-shot audit restores nominal
coverage under domain shift, wavelet error, and sparse acquisition on the
full Marmousi-2 line while flagging physics mismatch through its coverage
ceiling. Budget-matched baselines show that neither fixed-geometry
training nor explicit source conditioning attains any of these
properties. The complete experimental matrix runs in roughly 90 hours on
a single laptop; all scripts and checkpoints will be released and
archived on Zenodo.

\section*{Supplementary material}
Supplementary Figs.~S1--S6 (salt-benchmark validation, Marmousi FWI
stages, fine-tuning, and the elastic-mismatch experiment) and
Supplementary Tables~S1--S6 (complete crop-scale sweep and audit values,
exact per-family results, reproducibility parameters, baseline
hyperparameters, and the audit sensitivity analysis) accompany this
article as a single PDF.

\section*{Code and data availability}
All code, trained checkpoints, configuration files, and scripts (corpus
generation, classical chains, training, calibration, audit, sweeps, and
figures) will be released upon acceptance and archived on Zenodo with a
version DOI. The Marmousi-2 model \citep{martin2006} is publicly
available. No proprietary data are used.

\section*{CRediT authorship contribution statement}
\textbf{Deepak Kumar}: Conceptualization, Methodology, Software,
Investigation, Writing -- original draft, Writing -- review \& editing.
\textbf{Jayant Nath Tripathi}: Conceptualization, Supervision, Writing --
review \& editing.
\textbf{Laxmidhar Behera}: Supervision, Validation, Writing -- review \&
editing.

\section*{Declaration of competing interest}
The authors declare no competing financial interests.

\clearpage

\clearpage

\begin{figure}[p]
\centering
\includegraphics[width=\textwidth]{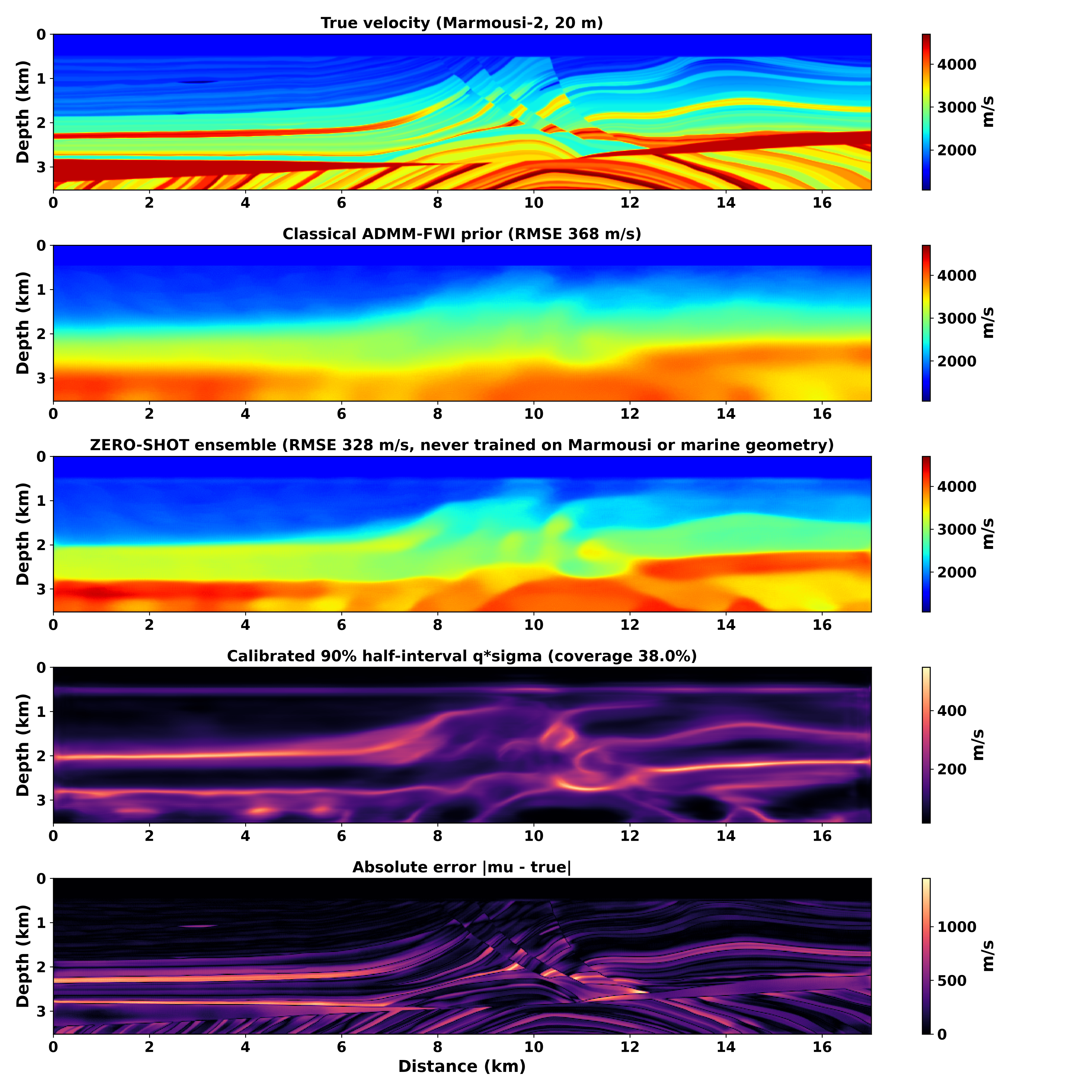}
\caption{Physics-mismatch experiment: elastic Marmousi-2 data inverted
with acoustic operators. The classical chain is fully neutralized, yet
the zero-shot ensemble still improves the prior; the audit detects and
repairs the induced miscalibration (Section~\ref{sec:marmousi}).}
\label{fig:elastic}
\end{figure}

\begin{figure}[p]
\centering
\includegraphics[width=\textwidth]{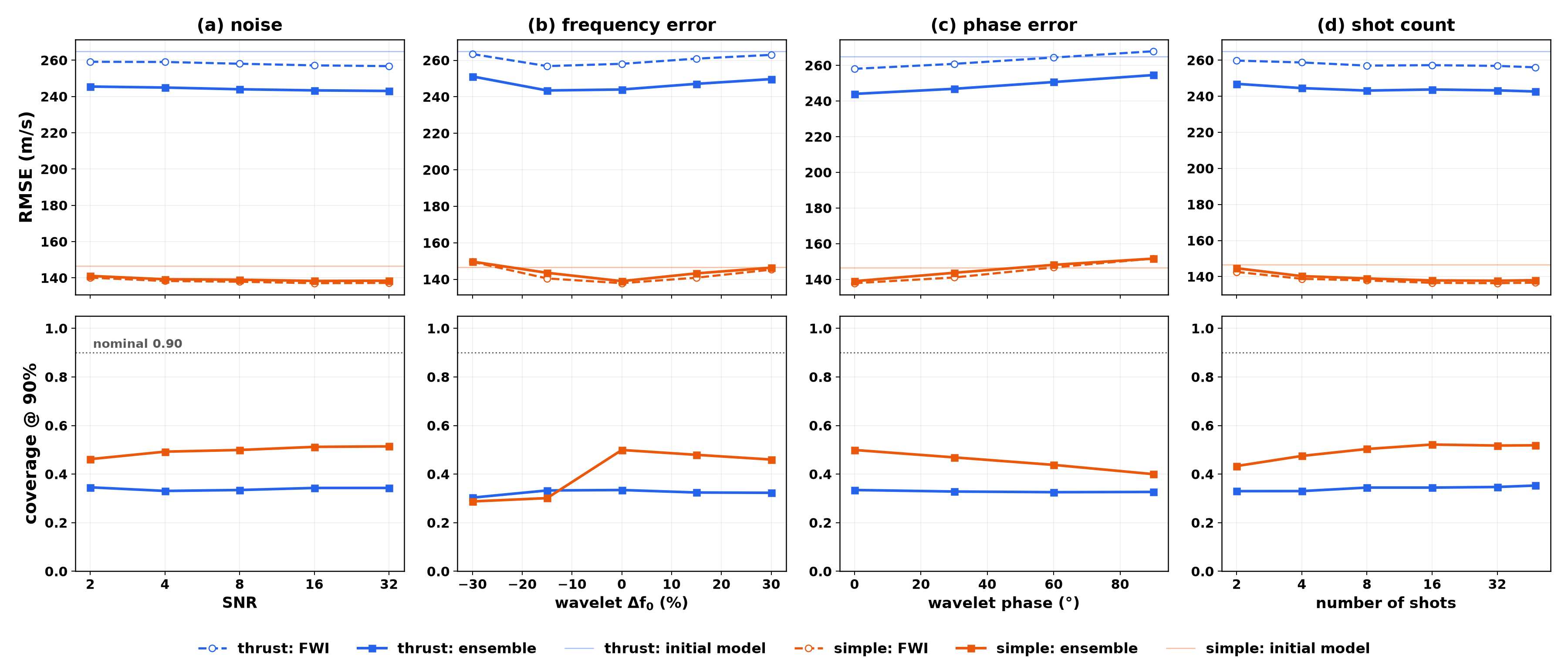}
\caption{Crop-scale sweeps on two 4$\times$2\,km Marmousi crops: RMSE and
coverage across the noise, wavelet-frequency, wavelet-phase, and
shot-count axes.}
\label{fig:crops}
\end{figure}

\clearpage
\setcounter{figure}{0}
\renewcommand{\thefigure}{S\arabic{figure}}
\setcounter{table}{0}
\renewcommand{\thetable}{S\arabic{table}}

\section*{Supplementary material}

\section*{Supplementary figures}

\begin{figure}[h!]
\centering
\includegraphics[width=\textwidth]{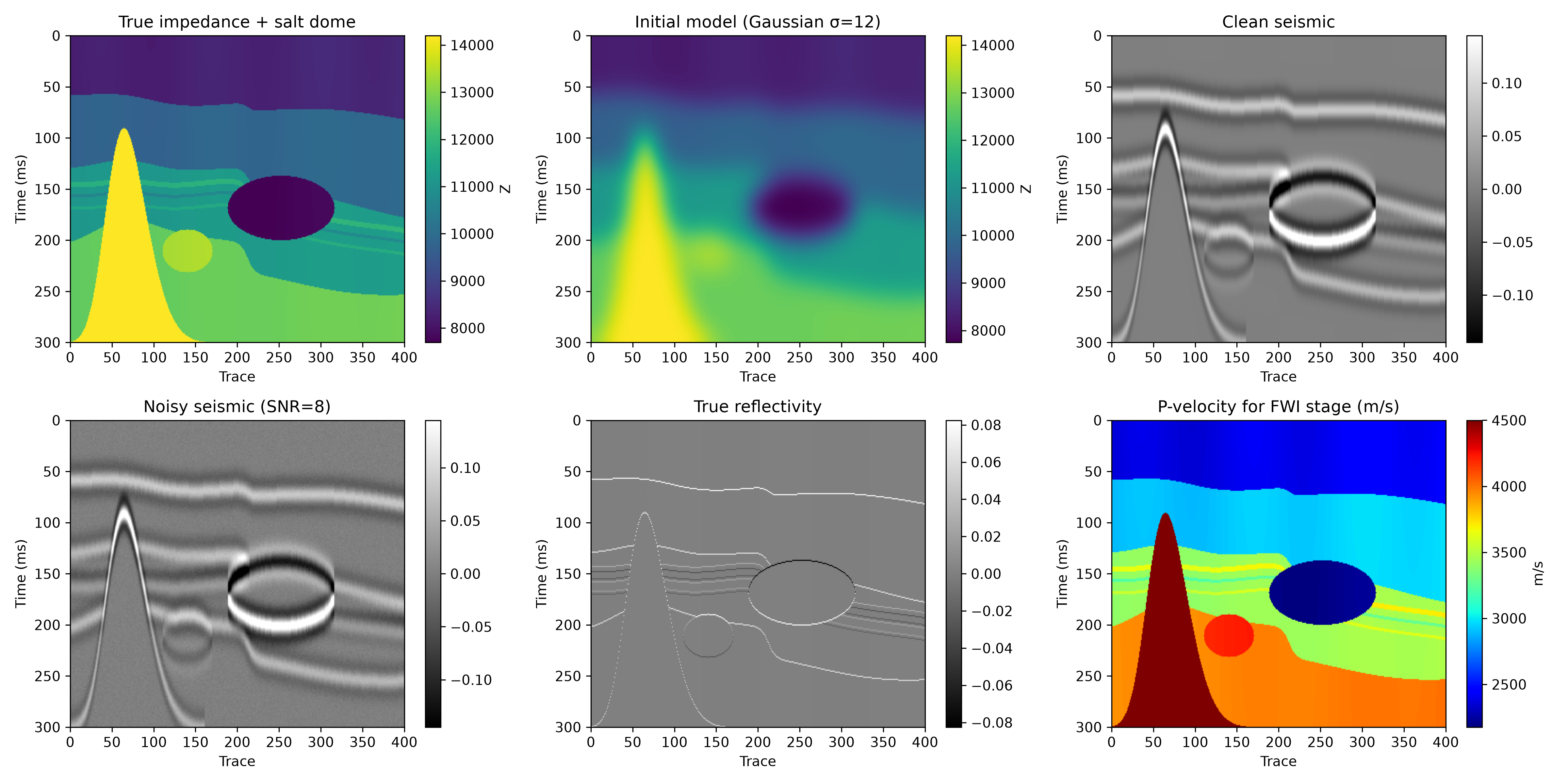}
\caption{Salt-diapir benchmark used to validate the classical chain before
corpus generation: the layered benchmark model of our earlier study with
an added salt body, its smoothed initial model, and the acquisition.}
\end{figure}

\begin{figure}[h!]
\centering
\includegraphics[width=\textwidth]{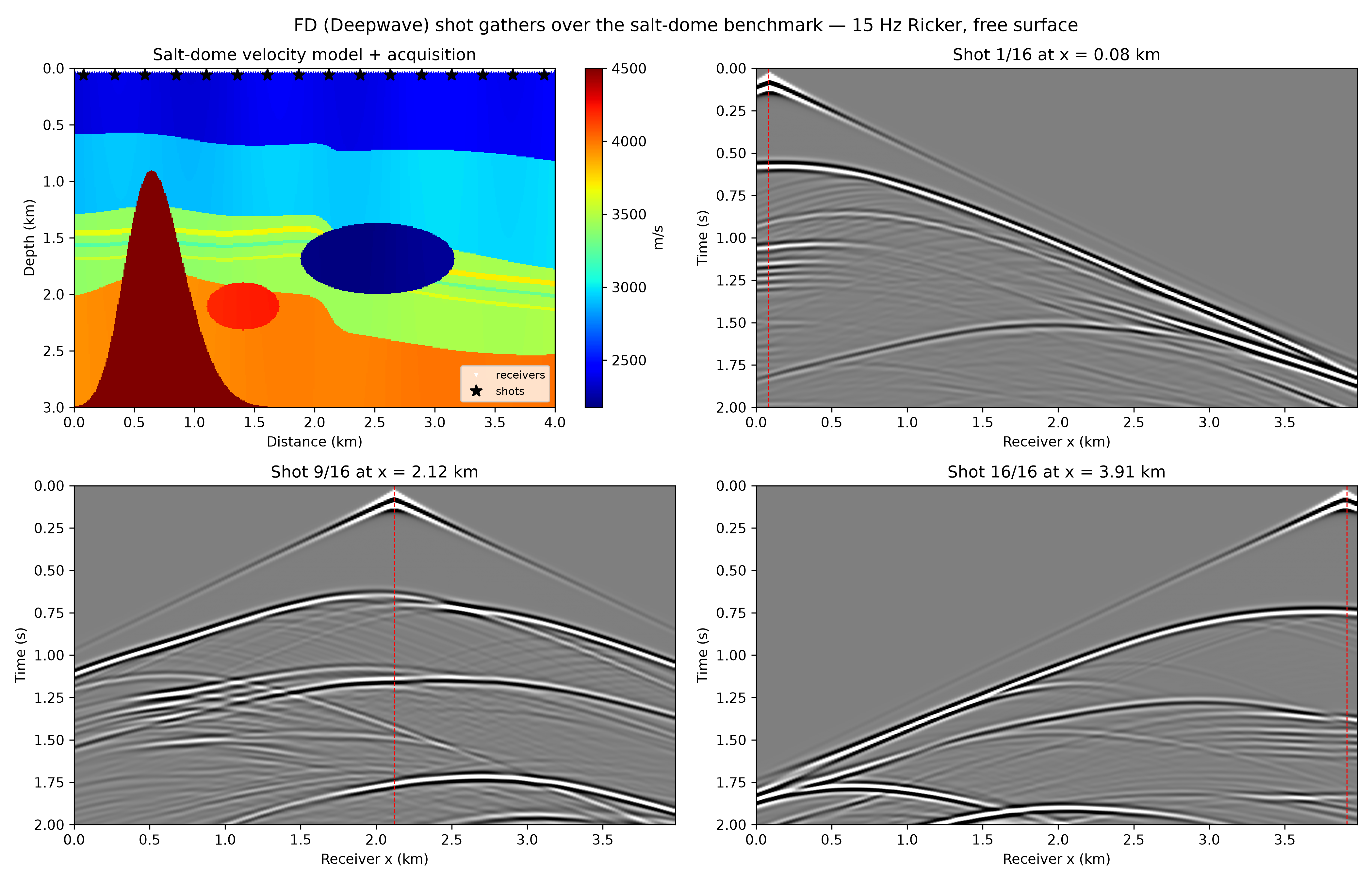}
\caption{Finite-difference data for the salt benchmark: acquisition layout
and three representative shot gathers used in the pilot validation of the
multiscale FWI and ADMM implementations.}
\end{figure}

\begin{figure}[h!]
\centering
\includegraphics[width=\textwidth]{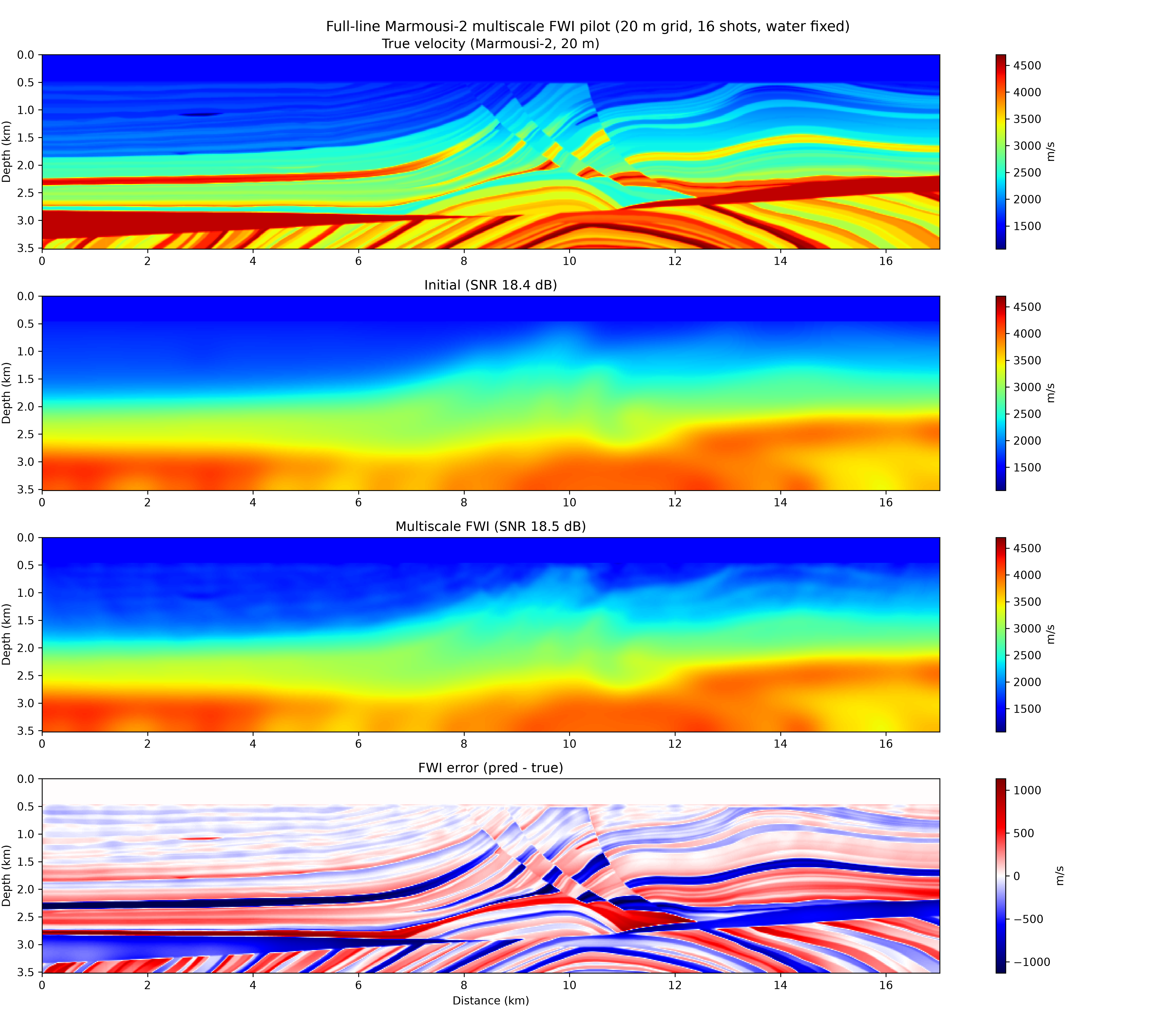}
\caption{Multiscale FWI pilot on the full Marmousi-2 line (clean data):
true model, initial model, inverted model after the 3--12\,Hz continuation,
and convergence history. This chain provides the classical stages that the
ADMM refinement and the learned ensemble build upon.}
\end{figure}

\begin{figure}[h!]
\centering
\includegraphics[width=\textwidth]{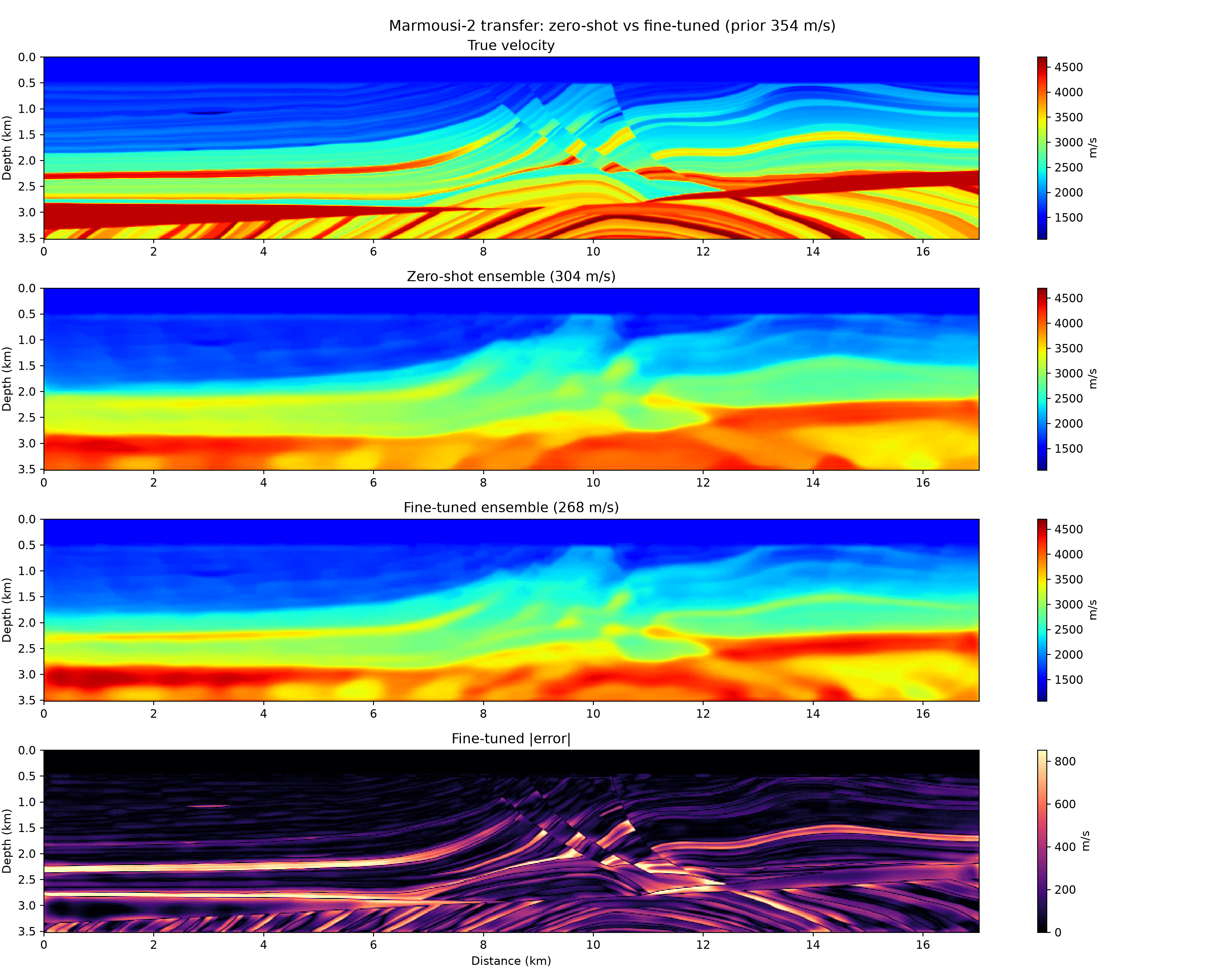}
\caption{Fine-tuning experiment on Marmousi-2: transfer-trained ensemble
(268\,m/s) compared with the zero-shot result (303\,m/s). Gains saturate
because the residual formulation inherits the reach of the classical
prior.}
\end{figure}

\begin{figure}[h!]
\centering
\includegraphics[width=\textwidth]{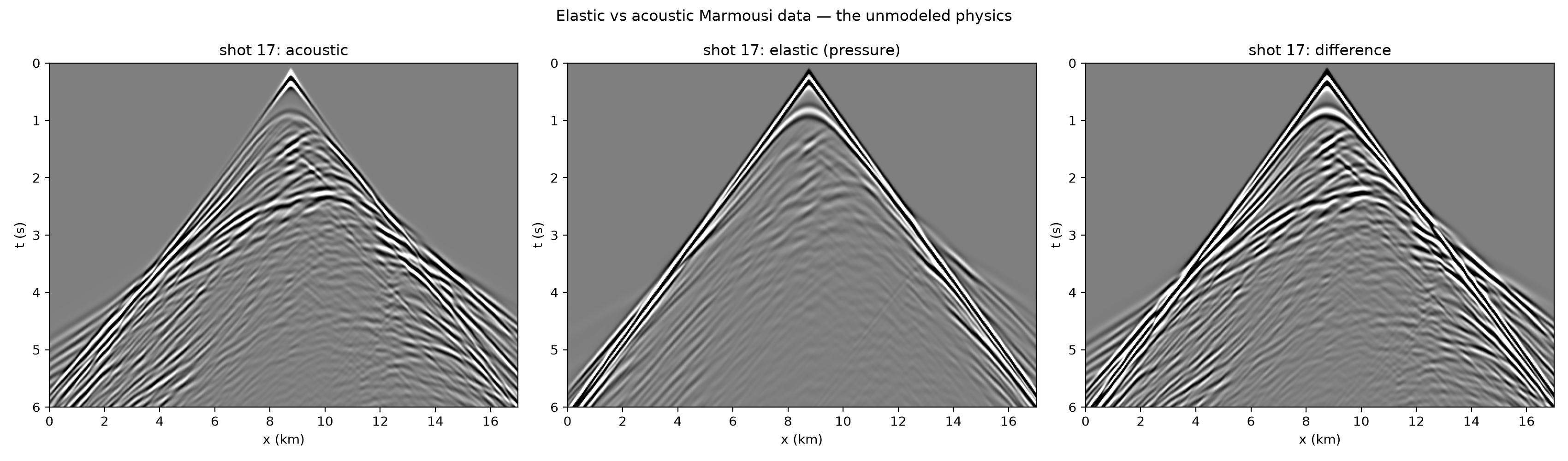}
\caption{Elastic versus acoustic finite-difference data for the
physics-mismatch experiment: pressure gathers from the elastic
simulation compared with their acoustic counterparts, illustrating the
amplitude and mode-conversion differences the acoustic operators cannot
explain.}
\end{figure}

\begin{figure}[h!]
\centering
\includegraphics[width=\textwidth]{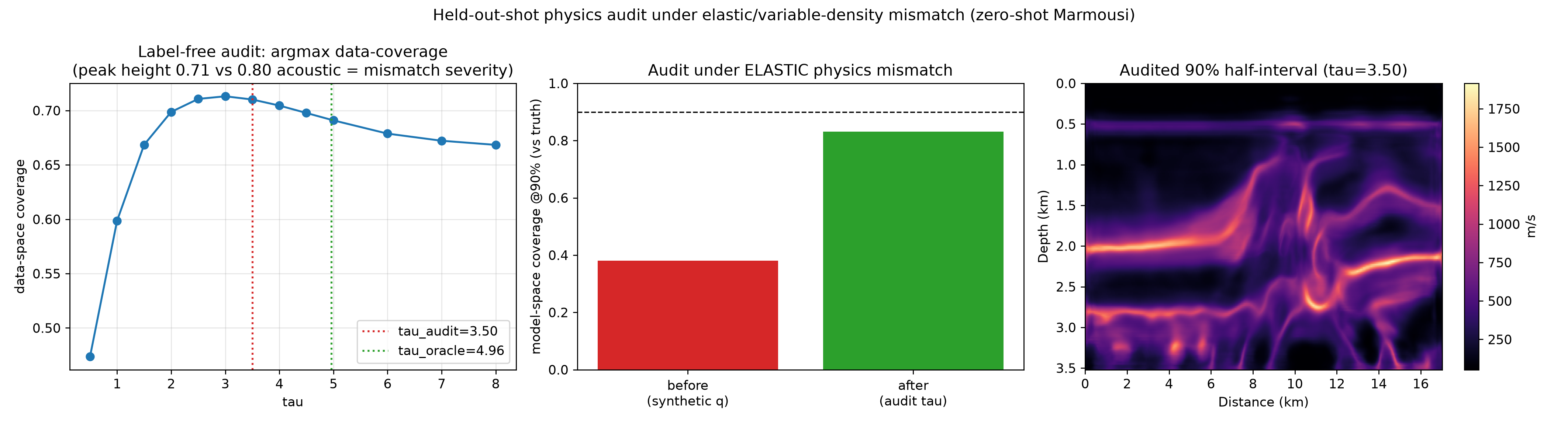}
\caption{Held-out-shot audit under elastic--acoustic physics mismatch:
the data-space coverage curve peaks lower (0.71 against the 0.80
matched-physics ceiling), flagging the mismatch, while its location still
repairs model-space coverage from 0.38 to 0.83.}
\end{figure}

\clearpage
\section*{Supplementary tables}

\begin{table}[h!]
\centering
\caption{Complete crop-scale robustness sweep (36 conditions; two
4$\times$2\,km Marmousi-2 crops at 10\,m). RMSE in m/s over sub-water
cells; $\bar\sigma$ is the mean predictive standard deviation; coverage is
evaluated at nominal 0.90 with the synthetic conformal quantile.}
\small
\begin{tabular}{@{}llrrrrrr@{}}
\toprule
Crop & Condition & Init & FWI & ADMM & Ensemble & $\bar\sigma$ &
Cov.@90\\
\midrule
thrust & SNR 2 & 265 & 259 & 259 & 245 & 92 & 0.346 \\
thrust & SNR 4 & 265 & 259 & 259 & 245 & 92 & 0.331 \\
thrust & SNR 8 & 265 & 258 & 258 & 244 & 92 & 0.335 \\
thrust & SNR 16 & 265 & 257 & 257 & 243 & 92 & 0.343 \\
thrust & SNR 32 & 265 & 257 & 257 & 243 & 92 & 0.343 \\
thrust & $\Delta f_0=-30\%$ & 265 & 263 & 264 & 251 & 89 & 0.303 \\
thrust & $\Delta f_0=-15\%$ & 265 & 257 & 257 & 243 & 90 & 0.333 \\
thrust & phase $30^\circ$ & 265 & 261 & 261 & 247 & 90 & 0.328 \\
thrust & phase $60^\circ$ & 265 & 264 & 264 & 251 & 90 & 0.326 \\
thrust & phase $90^\circ$ & 265 & 268 & 268 & 255 & 90 & 0.327 \\
thrust & $\Delta f_0=+15\%$ & 265 & 261 & 261 & 247 & 90 & 0.324 \\
thrust & $\Delta f_0=+30\%$ & 265 & 263 & 263 & 250 & 91 & 0.323 \\
thrust & 2 shots & 265 & 260 & 260 & 247 & 91 & 0.330 \\
thrust & 4 shots & 265 & 259 & 259 & 245 & 91 & 0.330 \\
thrust & 8 shots & 265 & 257 & 257 & 243 & 92 & 0.345 \\
thrust & 16 shots & 265 & 257 & 257 & 244 & 93 & 0.345 \\
thrust & 32 shots & 265 & 257 & 257 & 243 & 93 & 0.347 \\
thrust & 48 shots & 265 & 256 & 256 & 243 & 94 & 0.353 \\
simple & SNR 2 & 147 & 140 & 140 & 141 & 80 & 0.462 \\
simple & SNR 4 & 147 & 138 & 138 & 139 & 81 & 0.492 \\
simple & SNR 8 & 147 & 138 & 137 & 139 & 81 & 0.499 \\
simple & SNR 16 & 147 & 137 & 137 & 138 & 82 & 0.512 \\
simple & SNR 32 & 147 & 137 & 137 & 138 & 82 & 0.514 \\
simple & $\Delta f_0=-30\%$ & 147 & 150 & 149 & 150 & 62 & 0.288 \\
simple & $\Delta f_0=-15\%$ & 147 & 140 & 140 & 144 & 65 & 0.301 \\
simple & phase $30^\circ$ & 147 & 141 & 141 & 144 & 72 & 0.469 \\
simple & phase $60^\circ$ & 147 & 147 & 146 & 148 & 72 & 0.438 \\
simple & phase $90^\circ$ & 147 & 152 & 151 & 152 & 71 & 0.400 \\
simple & $\Delta f_0=+15\%$ & 147 & 141 & 141 & 143 & 73 & 0.479 \\
simple & $\Delta f_0=+30\%$ & 147 & 145 & 145 & 146 & 74 & 0.460 \\
simple & 2 shots & 147 & 143 & 142 & 145 & 75 & 0.433 \\
simple & 4 shots & 147 & 139 & 138 & 140 & 78 & 0.475 \\
simple & 8 shots & 147 & 138 & 137 & 139 & 81 & 0.503 \\
simple & 16 shots & 147 & 137 & 136 & 138 & 82 & 0.522 \\
simple & 32 shots & 147 & 136 & 136 & 138 & 82 & 0.518 \\
simple & 48 shots & 147 & 137 & 136 & 138 & 82 & 0.519 \\
\bottomrule
\end{tabular}
\end{table}

\begin{table}[h!]
\centering
\caption{Crop-scale audit spine (12 conditions). On short-offset crops the
$C(\tau)$ plateau makes $\tau$ unidentifiable and the plateau-right rule
runs to the grid edge ($\tau=12$), so repairs overshoot conservatively;
peak height still ranks mismatch severity (compare the wavelet conditions
with the reference).}
\small
\begin{tabular}{@{}llrrrrr@{}}
\toprule
Crop & Condition & $\tau_{\mathrm{audit}}$ & $\tau_{\mathrm{oracle}}$ &
Peak & Cov.\ before & Cov.\ after\\
\midrule
thrust & ref & 12.0 & 3.80 & 0.753 & 0.335 & 0.993 \\
thrust & snr2 & 12.0 & 3.79 & 0.799 & 0.346 & 0.993 \\
thrust & nsh2 & 12.0 & 3.89 & 0.837 & 0.330 & 0.993 \\
thrust & df-30 & 12.0 & 4.20 & 0.715 & 0.303 & 0.993 \\
thrust & df+30 & 12.0 & 4.02 & 0.731 & 0.323 & 0.992 \\
thrust & ph90 & 12.0 & 4.09 & 0.735 & 0.327 & 0.991 \\
simple & ref & 12.0 & 2.30 & 0.757 & 0.499 & 1.000 \\
simple & snr2 & 12.0 & 2.40 & 0.800 & 0.462 & 1.000 \\
simple & nsh2 & 12.0 & 2.67 & 0.811 & 0.433 & 1.000 \\
simple & df-30 & 12.0 & 4.09 & 0.693 & 0.288 & 0.997 \\
simple & df+30 & 12.0 & 2.58 & 0.714 & 0.460 & 1.000 \\
simple & ph90 & 12.0 & 2.75 & 0.723 & 0.400 & 1.000 \\
\bottomrule
\end{tabular}
\end{table}

\clearpage
\begin{table}[h!]
\centering
\caption{Exact per-family results on the 200-instance synthetic test set:
classical prior and ensemble RMSE (m/s), physics-conditioned conformal
coverage at nominal 0.90, and the no-augmentation ablation ensemble
(trained on a single fixed acquisition, evaluated on the same augmented
test set).}
\small
\begin{tabular}{@{}lrrrr@{}}
\toprule
Family & Prior & Ensemble & Cov.@90 (PC) & No-aug ensemble\\
\midrule
A (seen) & 229 & 144 & 0.890 & 152 \\
B (seen) & 204 & 122 & 0.898 & 131 \\
C (seen) & 184 & 111 & 0.918 & 115 \\
D (unseen) & 204 & 130 & 0.898 & 138 \\
E (unseen) & 194 & 122 & 0.900 & 128 \\
F (unseen) & 217 & 137 & 0.895 & 146 \\
\bottomrule
\end{tabular}
\end{table}

\begin{table}[h!]
\centering
\caption{Reproducibility parameters. Left: synthetic-corpus generator
distributions (all draws per instance from a generator seeded by
$10^4{+}\mathrm{index}$). Right: numerical and optimization parameters of
the classical chains. FWI steps are max-normalized preconditioned descent
(Eq.~3 of the main text) with step lengths in m/s.}
\footnotesize
\resizebox{\textwidth}{!}{%
\begin{tabular}{@{}ll@{}}
\toprule
\multicolumn{2}{@{}l}{\textbf{Corpus generator (per instance, seeded by
$10^4{+}\mathrm{index}$)}}\\
\midrule
Grid & $128\times256$ cells at 10\,m\\
Layers & 3--6; velocities 1600--4400\,m/s\\
Folds & dip $-15$ to $25^\circ$; amplitude 0--10 cells\\
Faults & 0--2; throw 5--18 cells\\
Thin beds & 0--3; amplitude $\pm$150--450\,m/s\\
Lenses & 0--2; velocities 1800--4400\,m/s\\
Salt body & probability 0.3; velocity 4400--4600\,m/s\\
Initial-model smoothing & Gaussian, 8--16 cells\\
Source & Ricker, $f_0 \in$ 6--15\,Hz\\
Wavelet error & prob.\ 0.5: $\Delta f_0\in[-30,30]\%$ or phase
0--$90^\circ$\\
Noise & band-limited, SNR log-uniform in [2,\,32]\\
Splits & 600 train / 100 val / 100 cal / 200 test (E/F in 900--999)\\
\midrule
\multicolumn{2}{@{}l}{\textbf{Numerics and classical chains}}\\
\midrule
Corpus modeling & 2\,s at 1\,ms, FD order 4, PML [0,15,15,15]\\
Corpus prior & 6 low-band GD steps + RW-$\ell_1$/TV ADMM (4$\times$2),
$\rho{=}0.02$, $\mu{=}0.2$\\
Marmousi modeling & 20\,m grid, 6\,s at 2\,ms, FD order 8, free
surface\\
Marmousi FWI & bands 3/5/8/12\,Hz, 15 iterations each, steps
15/12/9/7\,m/s\\
Preconditioning & smoothing $\sigma{=}2$ cells; illumination
compensation $\epsilon{=}0.05$; water masked\\
Marmousi ADMM & 8 outer $\times$ 3 Adam (lr 8\,m/s), $\rho{=}0.02$,
$\mu{=}0.2$, $\varepsilon_{\mathrm{rw}}{=}5$, reweight every 2\\
Velocity bounds & clip to 1000--4800\,m/s; water cells reset\\
Ensemble training & 6 members, 100 epochs, Adam $5\times10^{-4}$,
cosine decay, batch 8, $\lambda{=}0.1$\\
Audit defaults & $M{=}12$ samples, 60\,m correlation, $\tau$ grid
0.5--8, $\delta{=}0.005$\\
\bottomrule
\end{tabular}
}
\end{table}

\begin{table}[h!]
\centering
\caption{Baseline hyperparameters (both trained on the identical 600
training instances with the same augmentation, output grid, validation
split, and best-validation checkpoint selection as the ensemble).}
\footnotesize
\resizebox{\textwidth}{!}{%
\begin{tabular}{@{}lll@{}}
\toprule
 & Fixed-geometry InvNet & Source-conditioned net\\
\midrule
Input & 16 shots $\times$ 126 rec $\times$ 500 t & per-shot
$128\times500$ resampled gathers (max 16)\\
Conditioning & none (fixed acquisition) & FiLM on [shot $x$, $f_0$,
phase, $n_{\mathrm{shots}}$, rec.\ density]\\
Parameters & 35M & 35.4M\\
Training & 80 epochs, Adam $3\times10^{-4}$, cosine, batch 8 &
identical\\
\bottomrule
\end{tabular}
}
\end{table}

\begin{table}[h!]
\centering
\caption{Audit sensitivity analysis on the full-line reference condition
(oracle $\tau=4.24$; raw coverage 0.415). Each row changes one
hyperparameter from the defaults. The repaired coverage stays within
$0.90\pm0.08$ for every variant despite $4\times$ changes in sample
count and shot budget; the correlation length is the most sensitive
knob, confirming it should be set in physical units near the dominant
wavelength scale. Peak height is stable across held-out-shot choices
($\pm0.002$) but depends on $M$ through the sharpness of the empirical
percentile band, so peak-height comparisons are meaningful only at a
fixed audit configuration.}
\footnotesize
\begin{tabular}{@{}lccc@{}}
\toprule
Variant & $\tau_{\mathrm{audit}}$ & Peak $\max C(\tau)$ &
Cov.\ after repair\\
\midrule
defaults ($M{=}12$, 60\,m, 4 shots, $\delta{=}0.005$) & 4.0 & 0.803 & 0.888 \\
$M=6$ samples & 5.0 & 0.675 & 0.926 \\
$M=24$ samples & 3.5 & 0.857 & 0.859 \\
30\,m correlation length & 3.0 & 0.809 & 0.824 \\
120\,m correlation length & 7.0 & 0.776 & 0.963 \\
2 held-out shots & 4.5 & 0.801 & 0.910 \\
alternate held-out shot positions & 4.0 & 0.805 & 0.888 \\
plateau tolerance $\delta=0.02$ & 6.0 & 0.803 & 0.949 \\
\bottomrule
\end{tabular}
\end{table}


\begin{thebibliography}{48}
\expandafter\ifx\csname natexlab\endcsname\relax\def\natexlab#1{#1}\fi
\providecommand{\url}[1]{\texttt{#1}}
\providecommand{\href}[2]{#2}
\providecommand{\path}[1]{#1}
\providecommand{\DOIprefix}{doi:}
\providecommand{\ArXivprefix}{arXiv:}
\providecommand{\URLprefix}{URL: }
\providecommand{\Pubmedprefix}{pmid:}
\providecommand{\doi}[1]{\href{http://dx.doi.org/#1}{\path{#1}}}
\providecommand{\Pubmed}[1]{\href{pmid:#1}{\path{#1}}}
\providecommand{\bibinfo}[2]{#2}
\ifx\xfnm\relax \def\xfnm[#1]{\unskip,\space#1}\fi
\bibitem[{Adler et~al.(2021)Adler, Araya-Polo and Poggio}]{adler2021}
\bibinfo{author}{Adler, A.}, \bibinfo{author}{Araya-Polo, M.},
  \bibinfo{author}{Poggio, T.}, \bibinfo{year}{2021}.
\newblock \bibinfo{title}{Deep learning for seismic inverse problems: Toward
  the acceleration of geophysical analysis workflows}.
\newblock \bibinfo{journal}{IEEE Signal Processing Magazine}
  \bibinfo{volume}{38}, \bibinfo{pages}{89--119}.
\newblock \DOIprefix\doi{10.1109/MSP.2020.3037429}.
\bibitem[{Aghamiry et~al.(2019)Aghamiry, Gholami and Operto}]{aghamiry2019}
\bibinfo{author}{Aghamiry, H.S.}, \bibinfo{author}{Gholami, A.},
  \bibinfo{author}{Operto, S.}, \bibinfo{year}{2019}.
\newblock \bibinfo{title}{Implementing bound constraints and total-variation
  regularization in extended full-waveform inversion with the alternating
  direction method of multiplier: application to large contrast media}.
\newblock \bibinfo{journal}{Geophysical Journal International}
  \bibinfo{volume}{218}, \bibinfo{pages}{855--872}.
\newblock \DOIprefix\doi{10.1093/gji/ggz189}.
\bibitem[{Angelopoulos and Bates(2023)}]{angelopoulos2023}
\bibinfo{author}{Angelopoulos, A.N.}, \bibinfo{author}{Bates, S.},
  \bibinfo{year}{2023}.
\newblock \bibinfo{title}{Conformal prediction: A gentle introduction}.
\newblock \bibinfo{journal}{Foundations and Trends in Machine Learning}
  \bibinfo{volume}{16}, \bibinfo{pages}{494--591}.
\newblock \DOIprefix\doi{10.1561/2200000101}.
\bibitem[{Angelopoulos et~al.(2024)Angelopoulos, Bates, Fisch, Lei and
  Schuster}]{angelopoulos2024}
\bibinfo{author}{Angelopoulos, A.N.}, \bibinfo{author}{Bates, S.},
  \bibinfo{author}{Fisch, A.}, \bibinfo{author}{Lei, L.},
  \bibinfo{author}{Schuster, T.}, \bibinfo{year}{2024}.
\newblock \bibinfo{title}{Conformal risk control}, in:
  \bibinfo{booktitle}{International Conference on Learning Representations}.
\newblock \bibinfo{note}{ArXiv:2208.02814}.
\bibitem[{Araya-Polo et~al.(2018)Araya-Polo, Jennings, Adler and
  Dahlke}]{araya2018}
\bibinfo{author}{Araya-Polo, M.}, \bibinfo{author}{Jennings, J.},
  \bibinfo{author}{Adler, A.}, \bibinfo{author}{Dahlke, T.},
  \bibinfo{year}{2018}.
\newblock \bibinfo{title}{Deep-learning tomography}.
\newblock \bibinfo{journal}{The Leading Edge} \bibinfo{volume}{37},
  \bibinfo{pages}{58--66}.
\newblock \DOIprefix\doi{10.1190/tle37010058.1}.
\bibitem[{Boyd et~al.(2011)Boyd, Parikh, Chu, Peleato and Eckstein}]{boyd2011}
\bibinfo{author}{Boyd, S.}, \bibinfo{author}{Parikh, N.}, \bibinfo{author}{Chu,
  E.}, \bibinfo{author}{Peleato, B.}, \bibinfo{author}{Eckstein, J.},
  \bibinfo{year}{2011}.
\newblock \bibinfo{title}{Distributed optimization and statistical learning via
  the alternating direction method of multipliers}.
\newblock \bibinfo{journal}{Foundations and Trends in Machine Learning}
  \bibinfo{volume}{3}, \bibinfo{pages}{1--122}.
\newblock \DOIprefix\doi{10.1561/2200000016}.
\bibitem[{Brossier et~al.(2009)Brossier, Operto and Virieux}]{brossier2009}
\bibinfo{author}{Brossier, R.}, \bibinfo{author}{Operto, S.},
  \bibinfo{author}{Virieux, J.}, \bibinfo{year}{2009}.
\newblock \bibinfo{title}{Seismic imaging of complex onshore structures by {2D}
  elastic frequency-domain full-waveform inversion}.
\newblock \bibinfo{journal}{Geophysics} \bibinfo{volume}{74},
  \bibinfo{pages}{WCC105--WCC118}.
\newblock \DOIprefix\doi{10.1190/1.3215771}.
\bibitem[{Bunks et~al.(1995)Bunks, Saleck, Zaleski and Chavent}]{bunks1995}
\bibinfo{author}{Bunks, C.}, \bibinfo{author}{Saleck, F.M.},
  \bibinfo{author}{Zaleski, S.}, \bibinfo{author}{Chavent, G.},
  \bibinfo{year}{1995}.
\newblock \bibinfo{title}{Multiscale seismic waveform inversion}.
\newblock \bibinfo{journal}{Geophysics} \bibinfo{volume}{60},
  \bibinfo{pages}{1457--1473}.
\newblock \DOIprefix\doi{10.1190/1.1443880}.
\bibitem[{Cand{\`e}s et~al.(2008)Cand{\`e}s, Wakin and Boyd}]{candes2008}
\bibinfo{author}{Cand{\`e}s, E.J.}, \bibinfo{author}{Wakin, M.B.},
  \bibinfo{author}{Boyd, S.P.}, \bibinfo{year}{2008}.
\newblock \bibinfo{title}{Enhancing sparsity by reweighted $\ell_1$
  minimization}.
\newblock \bibinfo{journal}{Journal of Fourier Analysis and Applications}
  \bibinfo{volume}{14}, \bibinfo{pages}{877--905}.
\newblock \DOIprefix\doi{10.1007/s00041-008-9045-x}.
\bibitem[{Deng et~al.(2022)Deng, Feng, Wang, Zhang, Jin, Feng, Zeng, Chen and
  Lin}]{deng2022}
\bibinfo{author}{Deng, C.}, \bibinfo{author}{Feng, S.}, \bibinfo{author}{Wang,
  H.}, \bibinfo{author}{Zhang, X.}, \bibinfo{author}{Jin, P.},
  \bibinfo{author}{Feng, Y.}, \bibinfo{author}{Zeng, Q.},
  \bibinfo{author}{Chen, Y.}, \bibinfo{author}{Lin, Y.}, \bibinfo{year}{2022}.
\newblock \bibinfo{title}{{OpenFWI}: Large-scale multi-structural benchmark
  datasets for full waveform inversion}, in: \bibinfo{booktitle}{Advances in
  Neural Information Processing Systems}, pp. \bibinfo{pages}{6007--6020}.
\newblock \bibinfo{note}{ArXiv:2111.02926}.
\bibitem[{Ely et~al.(2018)Ely, Malcolm and Poliannikov}]{ely2018}
\bibinfo{author}{Ely, G.}, \bibinfo{author}{Malcolm, A.},
  \bibinfo{author}{Poliannikov, O.V.}, \bibinfo{year}{2018}.
\newblock \bibinfo{title}{Assessing uncertainties in velocity models and images
  with a fast nonlinear uncertainty quantification method}.
\newblock \bibinfo{journal}{Geophysics} \bibinfo{volume}{83},
  \bibinfo{pages}{R63--R75}.
\newblock \DOIprefix\doi{10.1190/geo2017-0321.1}.
\bibitem[{Esser et~al.(2018)Esser, Guasch, van Leeuwen, Aravkin and
  Herrmann}]{esser2018}
\bibinfo{author}{Esser, E.}, \bibinfo{author}{Guasch, L.}, \bibinfo{author}{van
  Leeuwen, T.}, \bibinfo{author}{Aravkin, A.Y.}, \bibinfo{author}{Herrmann,
  F.J.}, \bibinfo{year}{2018}.
\newblock \bibinfo{title}{Total variation regularization strategies in
  full-waveform inversion}.
\newblock \bibinfo{journal}{SIAM Journal on Imaging Sciences}
  \bibinfo{volume}{11}, \bibinfo{pages}{376--406}.
\newblock \DOIprefix\doi{10.1137/17M111328X}.
\bibitem[{Fichtner and Simut{\.e}(2018)}]{fichtner2018}
\bibinfo{author}{Fichtner, A.}, \bibinfo{author}{Simut{\.e}, S.},
  \bibinfo{year}{2018}.
\newblock \bibinfo{title}{Hamiltonian {Monte Carlo} inversion of seismic
  sources in complex media}.
\newblock \bibinfo{journal}{Journal of Geophysical Research: Solid Earth}
  \bibinfo{volume}{123}, \bibinfo{pages}{2984--2999}.
\newblock \DOIprefix\doi{10.1002/2017JB015249}.
\bibitem[{Gal and Ghahramani(2016)}]{gal2016}
\bibinfo{author}{Gal, Y.}, \bibinfo{author}{Ghahramani, Z.},
  \bibinfo{year}{2016}.
\newblock \bibinfo{title}{Dropout as a {Bayesian} approximation: Representing
  model uncertainty in deep learning}, in: \bibinfo{booktitle}{International
  Conference on Machine Learning}, pp. \bibinfo{pages}{1050--1059}.
\newblock \bibinfo{note}{ArXiv:1506.02142}.
\bibitem[{Gebraad et~al.(2020)Gebraad, Boehm and Fichtner}]{gebraad2020}
\bibinfo{author}{Gebraad, L.}, \bibinfo{author}{Boehm, C.},
  \bibinfo{author}{Fichtner, A.}, \bibinfo{year}{2020}.
\newblock \bibinfo{title}{Bayesian elastic full-waveform inversion using
  {Hamiltonian Monte Carlo}}.
\newblock \bibinfo{journal}{Journal of Geophysical Research: Solid Earth}
  \bibinfo{volume}{125}, \bibinfo{pages}{e2019JB018428}.
\newblock \DOIprefix\doi{10.1029/2019JB018428}.
\bibitem[{Guo et~al.(2017)Guo, Pleiss, Sun and Weinberger}]{guo2017}
\bibinfo{author}{Guo, C.}, \bibinfo{author}{Pleiss, G.}, \bibinfo{author}{Sun,
  Y.}, \bibinfo{author}{Weinberger, K.Q.}, \bibinfo{year}{2017}.
\newblock \bibinfo{title}{On calibration of modern neural networks}, in:
  \bibinfo{booktitle}{International Conference on Machine Learning}, pp.
  \bibinfo{pages}{1321--1330}.
\newblock \bibinfo{note}{ArXiv:1706.04599}.
\bibitem[{Kazei et~al.(2021)Kazei, Ovcharenko, Plotnitskii, Peter, Zhang and
  Alkhalifah}]{kazei2021}
\bibinfo{author}{Kazei, V.}, \bibinfo{author}{Ovcharenko, O.},
  \bibinfo{author}{Plotnitskii, P.}, \bibinfo{author}{Peter, D.},
  \bibinfo{author}{Zhang, X.}, \bibinfo{author}{Alkhalifah, T.},
  \bibinfo{year}{2021}.
\newblock \bibinfo{title}{Mapping full seismic waveforms to vertical velocity
  profiles by deep learning}.
\newblock \bibinfo{journal}{Geophysics} \bibinfo{volume}{86},
  \bibinfo{pages}{R711--R721}.
\newblock \DOIprefix\doi{10.1190/geo2019-0473.1}.
\bibitem[{Kendall and Gal(2017)}]{kendall2017}
\bibinfo{author}{Kendall, A.}, \bibinfo{author}{Gal, Y.}, \bibinfo{year}{2017}.
\newblock \bibinfo{title}{What uncertainties do we need in {Bayesian} deep
  learning for computer vision?}, in: \bibinfo{booktitle}{Advances in Neural
  Information Processing Systems}.
\newblock \bibinfo{note}{ArXiv:1703.04977}.
\bibitem[{Kumar and Tripathi(2026)}]{kumar2026}
\bibinfo{author}{Kumar, D.}, \bibinfo{author}{Tripathi, J.N.},
  \bibinfo{year}{2026}.
\newblock \bibinfo{title}{{ADMM}-guided physics-informed deep learning for
  two-dimensional acoustic impedance inversion with reweighted $\ell_1$ sparse
  regularization}.
\newblock \bibinfo{journal}{IEEE Transactions on Geoscience and Remote Sensing}
  \bibinfo{note}{Under review; EarthArXiv preprint, doi:10.31223/X56J49}.
\bibitem[{Lakshminarayanan et~al.(2017)Lakshminarayanan, Pritzel and
  Blundell}]{lakshminarayanan2017}
\bibinfo{author}{Lakshminarayanan, B.}, \bibinfo{author}{Pritzel, A.},
  \bibinfo{author}{Blundell, C.}, \bibinfo{year}{2017}.
\newblock \bibinfo{title}{Simple and scalable predictive uncertainty estimation
  using deep ensembles}, in: \bibinfo{booktitle}{Advances in Neural Information
  Processing Systems}.
\newblock \bibinfo{note}{ArXiv:1612.01474}.
\bibitem[{van Leeuwen and Herrmann(2013)}]{vanleeuwen2013}
\bibinfo{author}{van Leeuwen, T.}, \bibinfo{author}{Herrmann, F.J.},
  \bibinfo{year}{2013}.
\newblock \bibinfo{title}{Mitigating local minima in full-waveform inversion by
  expanding the search space}.
\newblock \bibinfo{journal}{Geophysical Journal International}
  \bibinfo{volume}{195}, \bibinfo{pages}{661--667}.
\newblock \DOIprefix\doi{10.1093/gji/ggt258}.
\bibitem[{Martin et~al.(2006)Martin, Wiley and Marfurt}]{martin2006}
\bibinfo{author}{Martin, G.S.}, \bibinfo{author}{Wiley, R.},
  \bibinfo{author}{Marfurt, K.J.}, \bibinfo{year}{2006}.
\newblock \bibinfo{title}{Marmousi2: An elastic upgrade for {Marmousi}}.
\newblock \bibinfo{journal}{The Leading Edge} \bibinfo{volume}{25},
  \bibinfo{pages}{156--166}.
\newblock \DOIprefix\doi{10.1190/1.2172306}.
\bibitem[{Mosser et~al.(2020)Mosser, Dubrule and Blunt}]{mosser2020}
\bibinfo{author}{Mosser, L.}, \bibinfo{author}{Dubrule, O.},
  \bibinfo{author}{Blunt, M.J.}, \bibinfo{year}{2020}.
\newblock \bibinfo{title}{Stochastic seismic waveform inversion using
  generative adversarial networks as a geological prior}.
\newblock \bibinfo{journal}{Mathematical Geosciences} \bibinfo{volume}{52},
  \bibinfo{pages}{53--79}.
\newblock \DOIprefix\doi{10.1007/s11004-019-09832-6}.
\bibitem[{Osypov et~al.(2013)Osypov, Yang, Fournier, Ivanova, Bachrach, Yarman,
  You, Nichols and Woodward}]{osypov2013}
\bibinfo{author}{Osypov, K.}, \bibinfo{author}{Yang, Y.},
  \bibinfo{author}{Fournier, A.}, \bibinfo{author}{Ivanova, N.},
  \bibinfo{author}{Bachrach, R.}, \bibinfo{author}{Yarman, C.E.},
  \bibinfo{author}{You, Y.}, \bibinfo{author}{Nichols, D.},
  \bibinfo{author}{Woodward, M.}, \bibinfo{year}{2013}.
\newblock \bibinfo{title}{Model-uncertainty quantification in seismic
  tomography: method and applications}.
\newblock \bibinfo{journal}{Geophysical Prospecting} \bibinfo{volume}{61},
  \bibinfo{pages}{1114--1134}.
\newblock \DOIprefix\doi{10.1111/1365-2478.12058}.
\bibitem[{Ovcharenko et~al.(2019)Ovcharenko, Kazei, Kalita, Peter and
  Alkhalifah}]{ovcharenko2019}
\bibinfo{author}{Ovcharenko, O.}, \bibinfo{author}{Kazei, V.},
  \bibinfo{author}{Kalita, M.}, \bibinfo{author}{Peter, D.},
  \bibinfo{author}{Alkhalifah, T.}, \bibinfo{year}{2019}.
\newblock \bibinfo{title}{Deep learning for low-frequency extrapolation from
  multioffset seismic data}.
\newblock \bibinfo{journal}{Geophysics} \bibinfo{volume}{84},
  \bibinfo{pages}{R989--R1001}.
\newblock \DOIprefix\doi{10.1190/geo2018-0884.1}.
\bibitem[{Paszke et~al.(2019)Paszke, Gross, Massa, Lerer, Bradbury, Chanan,
  Killeen, Lin, Gimelshein, Antiga et~al.}]{paszke2019}
\bibinfo{author}{Paszke, A.}, \bibinfo{author}{Gross, S.},
  \bibinfo{author}{Massa, F.}, \bibinfo{author}{Lerer, A.},
  \bibinfo{author}{Bradbury, J.}, \bibinfo{author}{Chanan, G.},
  \bibinfo{author}{Killeen, T.}, \bibinfo{author}{Lin, Z.},
  \bibinfo{author}{Gimelshein, N.}, \bibinfo{author}{Antiga, L.}, et~al.,
  \bibinfo{year}{2019}.
\newblock \bibinfo{title}{{PyTorch}: An imperative style, high-performance deep
  learning library}, in: \bibinfo{booktitle}{Advances in Neural Information
  Processing Systems}.
\newblock \bibinfo{note}{ArXiv:1912.01703}.
\bibitem[{Peters and Herrmann(2017)}]{peters2017}
\bibinfo{author}{Peters, B.}, \bibinfo{author}{Herrmann, F.J.},
  \bibinfo{year}{2017}.
\newblock \bibinfo{title}{Constraints versus penalties for edge-preserving
  full-waveform inversion}.
\newblock \bibinfo{journal}{The Leading Edge} \bibinfo{volume}{36},
  \bibinfo{pages}{94--100}.
\newblock \DOIprefix\doi{10.1190/tle36010094.1}.
\bibitem[{Plessix(2006)}]{plessix2006}
\bibinfo{author}{Plessix, R.{\'E}.}, \bibinfo{year}{2006}.
\newblock \bibinfo{title}{A review of the adjoint-state method for computing
  the gradient of a functional with geophysical applications}.
\newblock \bibinfo{journal}{Geophysical Journal International}
  \bibinfo{volume}{167}, \bibinfo{pages}{495--503}.
\newblock \DOIprefix\doi{10.1111/j.1365-246X.2006.02978.x}.
\bibitem[{Pratt(1999)}]{pratt1999}
\bibinfo{author}{Pratt, R.G.}, \bibinfo{year}{1999}.
\newblock \bibinfo{title}{Seismic waveform inversion in the frequency domain,
  part 1: Theory and verification in a physical scale model}.
\newblock \bibinfo{journal}{Geophysics} \bibinfo{volume}{64},
  \bibinfo{pages}{888--901}.
\newblock \DOIprefix\doi{10.1190/1.1444597}.
\bibitem[{Raissi et~al.(2019)Raissi, Perdikaris and Karniadakis}]{raissi2019}
\bibinfo{author}{Raissi, M.}, \bibinfo{author}{Perdikaris, P.},
  \bibinfo{author}{Karniadakis, G.E.}, \bibinfo{year}{2019}.
\newblock \bibinfo{title}{Physics-informed neural networks: A deep learning
  framework for solving forward and inverse problems involving nonlinear
  partial differential equations}.
\newblock \bibinfo{journal}{Journal of Computational Physics}
  \bibinfo{volume}{378}, \bibinfo{pages}{686--707}.
\newblock \DOIprefix\doi{10.1016/j.jcp.2018.10.045}.
\bibitem[{Richardson(2018)}]{richardson2018}
\bibinfo{author}{Richardson, A.}, \bibinfo{year}{2018}.
\newblock \bibinfo{title}{Seismic full-waveform inversion using deep learning
  tools and techniques}.
\newblock \bibinfo{journal}{arXiv preprint arXiv:1801.07232} .
\bibitem[{Romano et~al.(2019)Romano, Patterson and Cand{\`e}s}]{romano2019}
\bibinfo{author}{Romano, Y.}, \bibinfo{author}{Patterson, E.},
  \bibinfo{author}{Cand{\`e}s, E.J.}, \bibinfo{year}{2019}.
\newblock \bibinfo{title}{Conformalized quantile regression}, in:
  \bibinfo{booktitle}{Advances in Neural Information Processing Systems}.
\newblock \bibinfo{note}{ArXiv:1905.03222}.
\bibitem[{Ronneberger et~al.(2015)Ronneberger, Fischer and
  Brox}]{ronneberger2015}
\bibinfo{author}{Ronneberger, O.}, \bibinfo{author}{Fischer, P.},
  \bibinfo{author}{Brox, T.}, \bibinfo{year}{2015}.
\newblock \bibinfo{title}{U-{Net}: Convolutional networks for biomedical image
  segmentation}, in: \bibinfo{booktitle}{Medical Image Computing and
  Computer-Assisted Intervention}, \bibinfo{publisher}{Springer}. pp.
  \bibinfo{pages}{234--241}.
\newblock \DOIprefix\doi{10.1007/978-3-319-24574-4_28}.
\bibitem[{Rudin et~al.(1992)Rudin, Osher and Fatemi}]{rudin1992}
\bibinfo{author}{Rudin, L.I.}, \bibinfo{author}{Osher, S.},
  \bibinfo{author}{Fatemi, E.}, \bibinfo{year}{1992}.
\newblock \bibinfo{title}{Nonlinear total variation based noise removal
  algorithms}.
\newblock \bibinfo{journal}{Physica D: Nonlinear Phenomena}
  \bibinfo{volume}{60}, \bibinfo{pages}{259--268}.
\newblock \DOIprefix\doi{10.1016/0167-2789(92)90242-F}.
\bibitem[{Sethian(1996)}]{sethian1996}
\bibinfo{author}{Sethian, J.A.}, \bibinfo{year}{1996}.
\newblock \bibinfo{title}{A fast marching level set method for monotonically
  advancing fronts}.
\newblock \bibinfo{journal}{Proceedings of the National Academy of Sciences}
  \bibinfo{volume}{93}, \bibinfo{pages}{1591--1595}.
\newblock \DOIprefix\doi{10.1073/pnas.93.4.1591}.
\bibitem[{Shafer and Vovk(2008)}]{shafer2008}
\bibinfo{author}{Shafer, G.}, \bibinfo{author}{Vovk, V.}, \bibinfo{year}{2008}.
\newblock \bibinfo{title}{A tutorial on conformal prediction}.
\newblock \bibinfo{journal}{Journal of Machine Learning Research}
  \bibinfo{volume}{9}, \bibinfo{pages}{371--421}.
\newblock \bibinfo{note}{ArXiv:0706.3188}.
\bibitem[{Sirgue and Pratt(2004)}]{sirgue2004}
\bibinfo{author}{Sirgue, L.}, \bibinfo{author}{Pratt, R.G.},
  \bibinfo{year}{2004}.
\newblock \bibinfo{title}{Efficient waveform inversion and imaging: A strategy
  for selecting temporal frequencies}.
\newblock \bibinfo{journal}{Geophysics} \bibinfo{volume}{69},
  \bibinfo{pages}{231--248}.
\newblock \DOIprefix\doi{10.1190/1.1649391}.
\bibitem[{Sun et~al.(2020)Sun, Niu, Innanen, Li and Trad}]{sun2020}
\bibinfo{author}{Sun, J.}, \bibinfo{author}{Niu, Z.}, \bibinfo{author}{Innanen,
  K.A.}, \bibinfo{author}{Li, J.}, \bibinfo{author}{Trad, D.O.},
  \bibinfo{year}{2020}.
\newblock \bibinfo{title}{A theory-guided deep-learning formulation and
  optimization of seismic waveform inversion}.
\newblock \bibinfo{journal}{Geophysics} \bibinfo{volume}{85},
  \bibinfo{pages}{R87--R99}.
\newblock \DOIprefix\doi{10.1190/geo2019-0138.1}.
\bibitem[{Tarantola(1984)}]{tarantola1984}
\bibinfo{author}{Tarantola, A.}, \bibinfo{year}{1984}.
\newblock \bibinfo{title}{Inversion of seismic reflection data in the acoustic
  approximation}.
\newblock \bibinfo{journal}{Geophysics} \bibinfo{volume}{49},
  \bibinfo{pages}{1259--1266}.
\newblock \DOIprefix\doi{10.1190/1.1441754}.
\bibitem[{Virieux and Operto(2009)}]{virieux2009}
\bibinfo{author}{Virieux, J.}, \bibinfo{author}{Operto, S.},
  \bibinfo{year}{2009}.
\newblock \bibinfo{title}{An overview of full-waveform inversion in exploration
  geophysics}.
\newblock \bibinfo{journal}{Geophysics} \bibinfo{volume}{74},
  \bibinfo{pages}{WCC1--WCC26}.
\newblock \DOIprefix\doi{10.1190/1.3238367}.
\bibitem[{Vovk(2013)}]{vovk2012}
\bibinfo{author}{Vovk, V.}, \bibinfo{year}{2013}.
\newblock \bibinfo{title}{Conditional validity of inductive conformal
  predictors}.
\newblock \bibinfo{journal}{Machine Learning} \bibinfo{volume}{92},
  \bibinfo{pages}{349--376}.
\newblock \DOIprefix\doi{10.1007/s10994-013-5355-6}.
\bibitem[{Vovk et~al.(2005)Vovk, Gammerman and Shafer}]{vovk2005}
\bibinfo{author}{Vovk, V.}, \bibinfo{author}{Gammerman, A.},
  \bibinfo{author}{Shafer, G.}, \bibinfo{year}{2005}.
\newblock \bibinfo{title}{Algorithmic Learning in a Random World}.
\newblock \bibinfo{publisher}{Springer}, \bibinfo{address}{New York}.
\newblock \DOIprefix\doi{10.1007/b106715}.
\bibitem[{bin Waheed et~al.(2021)bin Waheed, Haghighat, Alkhalifah, Song and
  Hao}]{waheed2021}
\bibinfo{author}{bin Waheed, U.}, \bibinfo{author}{Haghighat, E.},
  \bibinfo{author}{Alkhalifah, T.}, \bibinfo{author}{Song, C.},
  \bibinfo{author}{Hao, Q.}, \bibinfo{year}{2021}.
\newblock \bibinfo{title}{{PINNeik}: Eikonal solution using physics-informed
  neural networks}.
\newblock \bibinfo{journal}{Computers \& Geosciences} \bibinfo{volume}{155},
  \bibinfo{pages}{104833}.
\newblock \DOIprefix\doi{10.1016/j.cageo.2021.104833}.
\bibitem[{Warner and Guasch(2016)}]{warner2016}
\bibinfo{author}{Warner, M.}, \bibinfo{author}{Guasch, L.},
  \bibinfo{year}{2016}.
\newblock \bibinfo{title}{Adaptive waveform inversion: Theory}.
\newblock \bibinfo{journal}{Geophysics} \bibinfo{volume}{81},
  \bibinfo{pages}{R429--R445}.
\newblock \DOIprefix\doi{10.1190/geo2015-0387.1}.
\bibitem[{Wu and Lin(2020)}]{wu2019}
\bibinfo{author}{Wu, Y.}, \bibinfo{author}{Lin, Y.}, \bibinfo{year}{2020}.
\newblock \bibinfo{title}{{InversionNet}: An efficient and accurate data-driven
  full waveform inversion}.
\newblock \bibinfo{journal}{IEEE Transactions on Computational Imaging}
  \bibinfo{volume}{6}, \bibinfo{pages}{419--433}.
\newblock \DOIprefix\doi{10.1109/TCI.2019.2956866}.
\bibitem[{Yang and Ma(2019)}]{yang2019}
\bibinfo{author}{Yang, F.}, \bibinfo{author}{Ma, J.}, \bibinfo{year}{2019}.
\newblock \bibinfo{title}{Deep-learning inversion: A next-generation seismic
  velocity model building method}.
\newblock \bibinfo{journal}{Geophysics} \bibinfo{volume}{84},
  \bibinfo{pages}{R583--R599}.
\newblock \DOIprefix\doi{10.1190/geo2018-0249.1}.
\bibitem[{Yin et~al.(2024)Yin, Orozco, Louboutin and Herrmann}]{yin2024wise}
\bibinfo{author}{Yin, Z.}, \bibinfo{author}{Orozco, R.},
  \bibinfo{author}{Louboutin, M.}, \bibinfo{author}{Herrmann, F.J.},
  \bibinfo{year}{2024}.
\newblock \bibinfo{title}{{WISE}: full-waveform variational inference via
  subsurface extensions}.
\newblock \bibinfo{journal}{Geophysics} \bibinfo{volume}{89},
  \bibinfo{pages}{A23--A28}.
\newblock \DOIprefix\doi{10.1190/geo2023-0744.1}.
\bibitem[{Zhu et~al.(2023)Zhu, Feng, Lin and Lu}]{zhu2023fourierdeeponet}
\bibinfo{author}{Zhu, M.}, \bibinfo{author}{Feng, S.}, \bibinfo{author}{Lin,
  Y.}, \bibinfo{author}{Lu, L.}, \bibinfo{year}{2023}.
\newblock \bibinfo{title}{{Fourier-DeepONet}: Fourier-enhanced deep operator
  networks for full waveform inversion with improved accuracy,
  generalizability, and robustness}.
\newblock \bibinfo{journal}{Computer Methods in Applied Mechanics and
  Engineering} \bibinfo{volume}{416}, \bibinfo{pages}{116300}.
\newblock \DOIprefix\doi{10.1016/j.cma.2023.116300}.

\end{thebibliography}
\end{document}